\documentclass[preprint,preprintnumbers,amsmath,amssymb,floatfix,nofootinbib]{revtex4}
\usepackage{epsf}
\usepackage{graphicx}
\usepackage{subfigure}

\newcommand{\beq}{\begin{equation}}
\newcommand{\eeq}{\end{equation}}

\def\fb{\rm fb}

\begin{document}

\preprint{~~PITT-PACC-1207}

\title{Resummation Effects in Vector-Boson and Higgs Associated Production}
\author{S.~Dawson$^{a}$}
\author{T. Han$^{b}$}
\author{W. K. Lai$^{b}$}
\author{A. K. Leibovich$^{b}$}
\author{I. Lewis$^{a}$}

\affiliation{
$^a$Department of Physics, Brookhaven National Laboratory,
Upton, NY 11973, USA\\
$^b$Pittsburgh Particle Physics Astrophysics and Cosmology Center (PITT PACC)\\
Department of Physics and Astronomy, University of Pittsburgh, Pittsburgh, PA 15260, USA
\vspace*{.5in}}

\date{\today}

\begin{abstract}
Fixed-order QCD radiative corrections to the vector-boson and Higgs associated production 
channels, $pp\rightarrow VH$ ($V=W^{\pm}, Z$),
at hadron colliders are well understood.  We combine higher order perturbative QCD calculations with soft-gluon resummation of both threshold logarithms and
logarithms which are important at low transverse momentum of
the $VH$ pair.  We study the effects of both types of logarithms on the scale 
dependence of the total cross section and on various kinematic distributions.
The  next-to-next-to-next-to-leading logarithmic
 (NNNLL) resummed total cross sections at the LHC are almost identical to the fixed-order perturbative next-to-next-to-leading order (NNLO) rates,  indicating the excellent convergence of the perturbative QCD series.
Resummation of the $VH$ transverse momentum ($p_T$) spectrum provides reliable results for small values of $p_T$
and suggests that implementing a jet-veto will significantly decrease the cross sections. 

\end{abstract}

\maketitle

\section{Introduction}
The recent discovery of a Higgs-like particle~\cite{atsem,cmssem} has brought our understanding of electroweak symmetry breaking to a deeper level.  Now  it is imperative to study the detailed properties of this particle in the hope of finding any hints for new physics beyond the Standard Model (SM).
An important Higgs production mechanism at hadron colliders is the associated production of  a Higgs boson and a vector boson, $VH\ (V=W^{\pm}, Z)$ \cite{Glashow:1978ab}.   
At the Tevatron, the process $q {\overline q}^\prime\rightarrow VH$ with the
decay of the vector boson to leptons and of the Higgs to the $b{\overline b}$ and $W^+W^-$ channels has provided important sensitivity to a light
Higgs boson \cite{Stange:1994bb,:2012cn}.  At the LHC, the production rate for associated $VH$  production is small, but with $\sim 30~\fb^{-1}$ a light Higgs in association with
a $W$ or $Z$ can potentially be observed in the boosted regime via $H\rightarrow b{\overline b}$ \cite{Butterworth:2008iy}.  Reliable predictions are essential for the observation and study of the $VVH$ couplings
 in this channel \cite{Dittmaier:2011ti,Dittmaier:2012vm}.

The rate for associated $VH$ production  is perturbatively known 
to next-to-next-to-leading order (NNLO), i.e.~${\cal O}(\alpha_s^2)$ \cite{Brein:2003wg,Brein:2011vx}. 
 At next-to-leading order (NLO), the QCD corrections are identical to those of the Drell-Yan process for an off-shell  gauge 
boson, $q {\overline q}^\prime
\rightarrow V^*$~\cite{Han:1991ia,Baer:1992vx,Ohnemus:1992bd}.  At NNLO, however, 
the $ZH$ process receives a small additional contribution from the $gg$ initial state, $gg\rightarrow ZH$ \cite{Brein:2003wg}. The NLO rates are available in the general purpose 
MCFM \cite{mcfm} program, while the total rate can
be found to NNLO using the VH@NNLO code \cite{Brein:2003wg}.

Infrared finite results in higher-order QCD processes occur due to a cancellation of virtual and real soft divergences. 
The fixed-order calculation is reliable providing all of the scales are of the same order of magnitude.
When the invariant mass $M_{VH}$ of the final state particles $WH$ or $ZH$ approaches the center-of-mass energy of the colliding partons, 
there is less phase space available for real emission.  While the infrared divergences will still cancel, large Sudakov logarithms will remain. 
 These logarithms can spoil the convergence of the perturbative series and need to be resummed to all orders for reliable results in this threshold region~\cite{Collins:1984kg}. 
Threshold
corrections involve terms of the form $\alpha_s^n {\log^{2n-1}(1-z)\over (1-z)}$, which are large when 
$z={M_{VH}^2/ {\hat s}}\sim 1$,
where ${\hat{s}}$ is the partonic center-of-mass (c.m.) energy-squared
~\cite{Sterman:1986aj,Catani:1996yz,Catani:1989ne,Sterman:2000pt,Becher:2007ty}.
Similarly, large logarithms  of the form 
$\alpha_s^n\log^{2n-1} \biggl({M_{VH}^2\over p_{T,VH}^2}\biggr)$
can also occur when the $VH$ system is produced with small transverse momentum $p_{T,VH}$~\cite{Bozzi:2005wk,Catani:2000vq}.
The techniques for resumming both types of logarithms to all orders are well known and the fixed order perturbative and resummed calculations 
can be consistently matched at intermediate
values of the kinematic variables.

We consider the process $pp
\rightarrow VH+X$ and present results from both the threshold resummation and the transverse momentum resummation of large logarithms
separately for LHC energies.  
Since the final state particles are color-singlets, both types of resummation can be straight-forwardly 
adopted from results in the literature for the Drell-Yan process~\cite{Becher:2007ty,Catani:2000vq,Bozzi:2005wk,Arnold:1990yk,Han:1991sa}. 
(We do not discuss the joint resummation of the 
logarithms \cite{Kulesza:2002rh}).  Section \ref{resum} contains a  brief review of
the resummation formalisms we apply.
Details are relegated to several appendices.   
Section \ref{sigres} presents results for the total cross section including the resummation of threshold logarithms
 and a discussion of the theoretical uncertainties, while Sections \ref{ptdts} and \ref{dts} 
contain some kinematic distributions 
resulting from the resummation of 
$p_{T,VH}$ and threshold logarithms, respectively.
Finally, Section \ref{conc} discusses the relevance of our results to searches
at the LHC.

\section{Resummation Formalism}
\label{resum}

In this section we briefly review the transverse momentum and threshold resummation formalism that we utilize
in deriving our numerical results.

\subsection{Transverse-Momentum Resummation}

The discussion of the transverse momentum resummation follows that of Grazzini {\it{et al.}}~\cite{Bozzi:2005wk}.
The hard scattering process under consideration is Higgs boson
 production in association with a vector boson in hadronic collisions 
\begin{equation}
AB\rightarrow V^*+X\rightarrow VH+X\, ,
\end{equation}
where $V=W^{\pm},Z$ and $X$ is the hadronic remnant of a collision.
We apply the well known impact-parameter space ($b$-space) resummation~\cite{Parisi:1979se,Collins:1981va} to the partonic cross section,
\begin{equation}
{d{\hat{\sigma}}_{VH}\over dM_{VH}^2dp_{T,VH}^2}
={d{\hat{\sigma}}_{VH}^{resum}\over dM_{VH}^2dp_{T,VH}^2}
+{d{\hat{\sigma}}_{VH}^{finite}\over dM_{VH}^2dp_{T,VH}^2}\, ,
\label{trans_1}
\end{equation}
where $p_{T,VH}$ is the transverse momentum of the $VH$
system and ${d{\hat{\sigma}}_{VH}^{resum}\over dM_{VH}^2dp_{T,VH}^2}$
contains the resummation of the 
$\log\biggl({M_{VH}^2\over p_{T,VH}^2}\biggr)$ enhanced terms.  
Since all the logarithmically enhanced terms are 
factored into the resummed piece, the remaining contribution ${d{\hat{\sigma}}_{VH}^{finite}\over dM_{VH}^2dp_{T,VH}^2}$
is finite as $p_{T,VH}\rightarrow 0$ and can be computed at fixed order in $\alpha_s$~\cite{Bozzi:2005wk}:
\begin{equation}
\left[{d{\hat{\sigma}}_{VH}^{finite}\over dM_{VH}^2
dp_{T,VH}^2}\right]_{f.o}
=\left[{d{\hat{\sigma}}_{VH}\over dM_{VH}^2dp_{T,VH}^2}
\right]_{f.o}-\left[{d{\hat{\sigma}}_{VH}^{resum}
\over dM_{VH}^2dp_{T,VH}^2}\right]_{f.o}\, ,
\label{finite.EQ}
\end{equation}
where the subscript $f.o.$ refers to a fixed order expansion.  In the low transverse momentum region, $p_{T,VH}\ll M_{VH}$, the 
resummed distribution is dominant, while in the high transverse momentum
 region, $p_{T,VH}\sim M_{VH}$, the perturbative expansion of the cross section dominates.  
Using Eq.~(\ref{finite.EQ}), the two regions can be consistently matched in the intermediate $p_{T,VH}$ region, maintaining theoretical accuracy.

To correctly account for momentum conservation, transverse momentum resummation is performed in impact-parameter space:
\begin{equation}
M_{VH}^2{d{\hat{\sigma}}_{VH}^{resum}\over dM_{VH}^2dp_{T,VH}^2}
={M_{VH}^2\over {\hat s}}
\int_0^\infty db 
{b\over 2}
J_0(bp_{T,VH})
W^{VH}(b,M_{VH}, {\hat s},\mu_r,\mu_f)\, ,
 \end{equation}
 where $J_0(x)$ is the $0^{th}$ order Bessel function and $\mu_r,\mu_f$ are the renormalization/factorization scales.  
 By performing a Mellin transformation\footnote{The Mellin transformation of a function $h(z)$ is defined as $h_N=\int^1_0 dz z^{N-1}h(z)$.} it is possible to factor the terms that are finite and logarithmically enhanced as 
$p_{T,VH}\rightarrow 0$:
 \begin{eqnarray}
 W_N^{VH}(b,M_{VH},\mu_r,\mu_f)&=&H_N^{VH}\left(M_{VH},\alpha_s(\mu_r),{M_{VH}\over \mu_r},{M_{VH} \over\mu_f},{M_{VH}\over Q}\right)\nonumber\\
&& \times\exp\biggr\{G_N\left(\alpha_s(\mu_r), L, {M_{VH}\over\mu_r},{M_{VH}\over Q}\right)\biggr\}\, ,
\label{WN.EQ}
 \end{eqnarray}
 where $L=\ln\biggl({Q^2 b^2/ b_0^2}\biggr)$ with  $b_0=2\exp(-\gamma_E)$, $H$ contains the finite hard scattering coefficients, and $G$ contains the process independent logarithmically enhanced terms.  
 Hence, all the terms that are divergent as $p_{T,VH}\rightarrow0$ 
are exponentiated into the function $G_N$, achieving the all-orders
 resummation. The split between the finite and logarithmically enhanced terms is somewhat arbitrary;
 that is, a finite shift in the invariant mass $M_{VH}$ can alter the separation:
\begin{eqnarray}
\log\biggl({M_{VH}^2\over p_{T,VH}^2}\biggr)
=\log\left({Q^2\over p_{T,VH}^2}\right)+\log\left({M_{VH}^2\over Q^2}\right).
\end{eqnarray}
The scale $Q$, termed the resummation scale, is introduced to parameterize this arbitrariness and is the same as that in Eq.~(\ref{WN.EQ}).  To keep the separation between the finite and logarithmically enhanced terms meaningful, the scale $Q$ has to be chosen to be close to $M_{VH}$.

As mentioned in the previous paragraph, all of the logarithmically enhanced contributions are contained in $G_N$. The divergent pieces can be reorganized such that $G_N$ is written as an expansion that is order-by-order smaller by $\alpha_s$~\cite{Bozzi:2005wk}:
 \begin{equation}
 G_N\left(\alpha_s,L,{M_{VH}\over\mu_r},{M_{VH}\over Q}\right)=Lg_N^1(\alpha_s L)+\sum_{n=2} \left(\frac{\alpha_s}{\pi}\right)^{n-2} g_N^n\left(\alpha_s L,{M_{VH}\over\mu_r},{M_{VH}\over Q}\right),
 \end{equation}
where $g^n_N=0$ for $\alpha_sL=0$ and $Lg_N^1$ contains the leading log (LL) terms $\alpha_s^n L^{n+1}$, $g_N^2$ contains the next-to-leading log (NLL) terms $\alpha_s^n L^n$, etc.  Since the large 
logarithms are associated with collinear and soft divergences from real radiation, the functions $g_N^i$ are only dependent on the initial state partons and are independent of the specific hard process under consideration.  Explicit expressions for the LL and NLL terms needed for $pp
\rightarrow VH +X$ are given in Appendix A.  

The resummed distribution is valid in the low 
$p_{T,VH}\ll M_{VH}$ region, while the perturbative expansion 
is valid in the high $p_{T,VH}\sim M_{VH}$ region.  However, as $Qb$ approaches zero the logarithm $L$ grows uncontrollably.  As a result, 
the resummed distribution makes an unacceptably
large contribution to the high $p_{T,VH}$ region.  
This problem can be solved via the replacement~\cite{Catani:1992ua} 
$L\rightarrow \tilde{L}=\log\biggl(
 {Q^2 b^2/ b_0^2}+1\biggr)$, such that $\tilde{L}\approx L$ for $Qb\gg 1$ and $\tilde{L}\approx 0$ for $Qb \ll 1$.  
Hence, using $\tilde{L}$, the resummed contribution maintains the correct 
dependence on the large logarithms at low $p_{T,VH}$ and does not make 
unwarranted contributions to the high $p_{T,VH}$ region.  
This replacement has the added benefit of reproducing the correct fixed order cross section once the transverse momentum is integrated~\cite{Bozzi:2005wk}.

The process-dependent function $H$ is finite as $p_{T,VH}\rightarrow 0$.  Hence, its Mellin transform $H_N$ does not contain any dependence on $b$ and can be computed as an
expansion in $\alpha_s$,
\begin{eqnarray}
&& H_N^{VH}\left(M_{VH},\alpha_s,\frac{M_{VH}}{\mu_r},\frac{M_{VH}}{\mu_r},\frac{M_{VH}}{Q}\right) 
\nonumber \\
&& =
\sigma_0(\alpha_s,M_{VH})\biggl\{ 1+\sum_{n=1}\left({\alpha_s\over \pi}\right)^n H_N^{VH(n)}\left(\frac{M_{VH}}{\mu_r},\frac{M_{VH}}{\mu_f},\frac{M_{VH}}{Q}\right)\biggr\} ,
\end{eqnarray}
where $\sigma_0$ is the Born-level partonic cross section for 
$q {\overline q}^\prime \rightarrow VH$.  At NLL accuracy, only the first hard coefficient $H_N^{VH(1)}$ is needed.  The value of this coefficient is given in Appendix A.

\label{sec_trans}

\subsection{Threshold resummation}
In the original approach
to threshold resummation~\cite{Sterman:1986aj,Catani:1989ne},
the resummation is performed after taking the Mellin 
transformation of the hadronic 
cross section~\cite{Magnea:1990qg,Korchemsky:1993uz}.
The Mellin-transformed hadronic cross 
section can then be factored into the product of the partonic cross section 
and the parton luminosity.  
The threshold logarithms for $VH$ production
 are  of the form $\ln(1-z)$, where $z=M_{VH}^2/\hat s$, 
and are contained in the partonic cross section.  
After resummation, an inverse-Mellin transformation is 
performed to obtain the physical cross section.  
This leads to a new divergence due to the presence of the Landau pole in $\alpha_s$.  Prescriptions for how to 
perform the inverse-Mellin transformation 
have been developed to remove this problem.
The resummation of
threshold logarithms for Drell-Yan production has been extensively 
studied~\cite{Bolzoni:2006ky,Mukherjee:2006uu,Ravindran:2007sv,Ravindran:2006bu}.

More recently, techniques using soft-collinear effective 
theory (SCET)~\cite{Bauer:2000ew,Bauer:2000yr,Bauer:2001yt,Beneke:2002ph} 
have been developed in which the resummation is performed 
in momentum space, obviating the need to go to Mellin space.  
This in turn removes the problem of the Landau pole.   In this paper, we will
 generalize the SCET resummation results of~\cite{Becher:2007ty} to the
case of $VH$ production.

The leading singular terms at threshold in the
hadronic differential cross section can be written as
\begin{equation}
\frac1{\tau\sigma_0}\frac{d\sigma}{dM_{VH}^2} 
= \int_\tau^1\frac{dz}z C(z,M_{VH},\mu_f){\cal L} \left(\frac\tau{z},\mu_f\right),\label{threshold_sigma}
\end{equation}
where $\tau = M_{VH}^2/s$ with $s$ the hadronic c.m.~energy-squared, ${\cal L}$ is the parton luminosity,
\begin{equation}
{\cal L}(y,\mu_f)
=\int_y^1 
{dx\over x} 
f_{q}(x,\mu_f) 
f_{{\overline q}^\prime}
\biggl({y\over x},\mu_f\biggr)+(q\leftrightarrow {\overline q}^\prime)
\, ,\label{dsigma_dM2}
\end{equation} 
 and $\sigma_0$ is the Born level partonic cross section for 
$q {\overline q}^\prime \rightarrow VH$ and is  defined such
 that $C(z,M_{VH},\mu_f) = \delta(1-z) + {\cal O}(\alpha_s)$. 
In the threshold region, $z\sim 1$, $C(z,M_{VH},\mu_f)$ can be factorized into 
a hard contribution and a soft contribution,
\begin{equation}
C(z,M_{VH},\mu_f) = {\cal H}(M_{VH}, \mu_f)
{\cal S}(M_{VH}(1-z),\mu_f).\label{SCET_factorization}
\end{equation}
The hard function ${\cal H}(M_{VH}, \mu_f)$ 
and soft function ${\cal S}(M_{VH}(1-z),\mu_f)$,
 evaluated at $\mu_f$, are obtained by 
renormalization group running from the hard scale $\mu_h\sim M_{VH}$ 
and soft scale $\mu_s\sim M_{VH}(1-\tau)$, respectively, 
to sum the threshold logarithms to all orders in $\alpha_s$.

The final result is found from that for the Drell-Yan process~\cite{Becher:2007ty} 
\begin{eqnarray}
C(z,M_{VH},\mu_f)&=&
\left|C_V(-M_{VH}^2,\mu_h)\right|^2U(M_{VH},\mu_h,\mu_s,\mu_f)\frac{z^{-\eta}}{(1-z)^{1-2\eta}}\nonumber \\
&&\times \tilde{s}_{DY}\left(
\ln\frac{M_{VH}^2(1-z)^2}{\mu_s^2z}+\partial_\eta,\mu_s\right)\frac{e^{-2\gamma_E\eta}}{\Gamma(2\eta)},\label{threshold_resum}
\end{eqnarray}
where $\eta = 2 a_\Gamma(\mu_s, \mu_f)$,  and $C_V$ and $\tilde{s}_{DY}$ 
are the  perturbatively calculable Wilson coefficient and soft Wilson loop
coefficient, respectively. Eq.~(\ref{dsigma_dM2}) with $C$ given by Eq.~(\ref{threshold_resum}) is defined only for $\eta>0$. For 
$\eta<0$, an analytic continuation is required.  
The analytic expressions for $a_\Gamma, C_V, {\tilde s}_{DY}$ and $U$ 
which are necessary for our numerical calculations are 
given in Appendix B.

Eq.~(\ref{threshold_resum}) is only valid in the threshold region $z\sim 1$. To obtain a formula valid 
for all values of $z$, we match the threshold-resummed result with the fixed-order result,
\begin{equation}
\biggl[\frac{d\sigma}{dM_{VH}^2}\biggr]_{matched}
=\biggl[\frac{d\sigma}{dM_{VH}^2}\biggr]_{threshold~resum}
-\biggl[\frac{d\sigma}{dM_{VH}^2}\biggr]_{threshold~f.o.}
+\biggl[\frac{d\sigma}{dM_{VH}^2}\biggr]_{f.o.}\,. \label{threshold_combine}
\end{equation}
Here $\biggl[\frac{d\sigma}{dM_{VH}^2}\biggr]_{threshold~resum}$ 
is the result obtained using the threshold resummation formula 
of Eq.~(\ref{threshold_resum}), 
$\biggl[\frac{d\sigma}{dM_{VH}^2}\biggr]_{f.o.}$ is the fixed-order 
perturbative result and 
$\biggl[\frac{d\sigma}{dM_{VH}^2}\biggr]_{threshold~f.o.}$ 
is obtained from the fixed-order result by 
keeping only the leading threshold singularity in $C$. 
The order of the  logarithmic approximation in 
the resummed result and the corresponding fixed-order results
used in the matching of Eq.~(\ref{threshold_combine}) are
 summarized in Table \ref{threshold_resummation_orders}.\footnote{
The equivalence of the sub-leading logarithms 
between the SCET approach and the standard QCD Mellin transform
approach has been studied in Ref.
\cite{Bonvini:2012yg}.}

\begin{table}[tp]
\caption{Approximation schemes for threshold resummation given a fixed order matched to a logarithmic approximation as in Eq.~(\ref{threshold_combine}).}
\label{threshold_resummation_orders}\centering
\begin{tabular}{cccccc}
\hline
Fixed order~~  & Log.~~ & Accuracy$\sim\alpha_s^nL^k$~~ & $\Gamma_{\textrm{cusp}}$~~ & $\gamma^V,\gamma^\phi$~~ & $C_V, \tilde{s}_{DY}$ \\\hline
LO                   &  NLL         &$2n-1\le k\le 2n$            & 2-loop          & 1-loop                 & tree-level           \\
NLO                  & NNLL         &$2n-3\le k\le 2n$            & 3-loop          & 2-loop                 & 1-loop               \\
NNLO                 & NNNLL        &$2n-5\le k\le 2n$            & 4-loop          & 3-loop                 & 2-loop                \\
\hline
\end{tabular}
\end{table}

\label{sec_thresh}

\section{Scale Dependence of the Cross Section}
\label{sigres}

In this section, we study  the scale dependence of
the total cross section for $VH$ production at the LHC,
beginning with the sensitivity of the resummed threshold distributions
to the hard, soft, and factorization scales.  Near the threshold, 
$\tau\equiv {M_{VH}^2/ s}\rightarrow 1$, the threshold logarithms are enhanced, leading to
potentially large scale violations.
The naive choice for the soft scale is $\mu_s\sim M_{VH}(1-\tau)$.  We
follow the prescription of Ref.~\cite{Becher:2007ty} to determine a sensible range of parameters
for the soft scale.  A low value of $\mu_s$ is found empirically from the scale
where the one-loop correction to $\tilde{s}_{DY}$ is minimal,
\begin{equation}
\mu_s^{(I)}={M_{VH}(1-\tau)\over 2\sqrt{1+100\tau}}\, .
\end{equation}
Alternatively, an upper scale for the soft variation can be chosen as
the value where the one-loop correction to $\tilde{s}_{DY}$ drops 
below $10\%$,
\begin{equation}
\mu_s^{(II)}={M_{VH}(1-\tau)\over 0.9+12\tau}
\, .
\end{equation}
Empirically, the forms of ${\mu_s^{(I,II)}\over M_{VH}}$ are insensitive to $M_{VH}$. Here and henceforth, we adopt the Higgs mass value 
\begin{equation}
M_{H} = 125\ {\rm GeV}.
\end{equation}

We investigate the numerical effects of the scale variation by plotting
the differential 
cross section of the threshold resummation of Eq.~(\ref{threshold_resum}) and varying
the soft, hard, and factorization scales.  It is customary to measure the size of QCD corrections by a 
$K$-factor ($K$) typically defined as the ratio of a higher order cross section to the lowest order cross section:
\begin{equation}
\frac{d\sigma}{dM_{VH}^2}\equiv K\frac{d\sigma}{dM_{VH}^2}\bigg|_{\textrm{LO}},
\label{eq:K}
\end{equation}
where $\frac{d\sigma}{dM_{VH}^2}$ is a distribution defined
 at higher order in QCD.

\begin{figure}
\begin{center}
\subfigure[]{
\includegraphics[width=0.48\textwidth,clip]{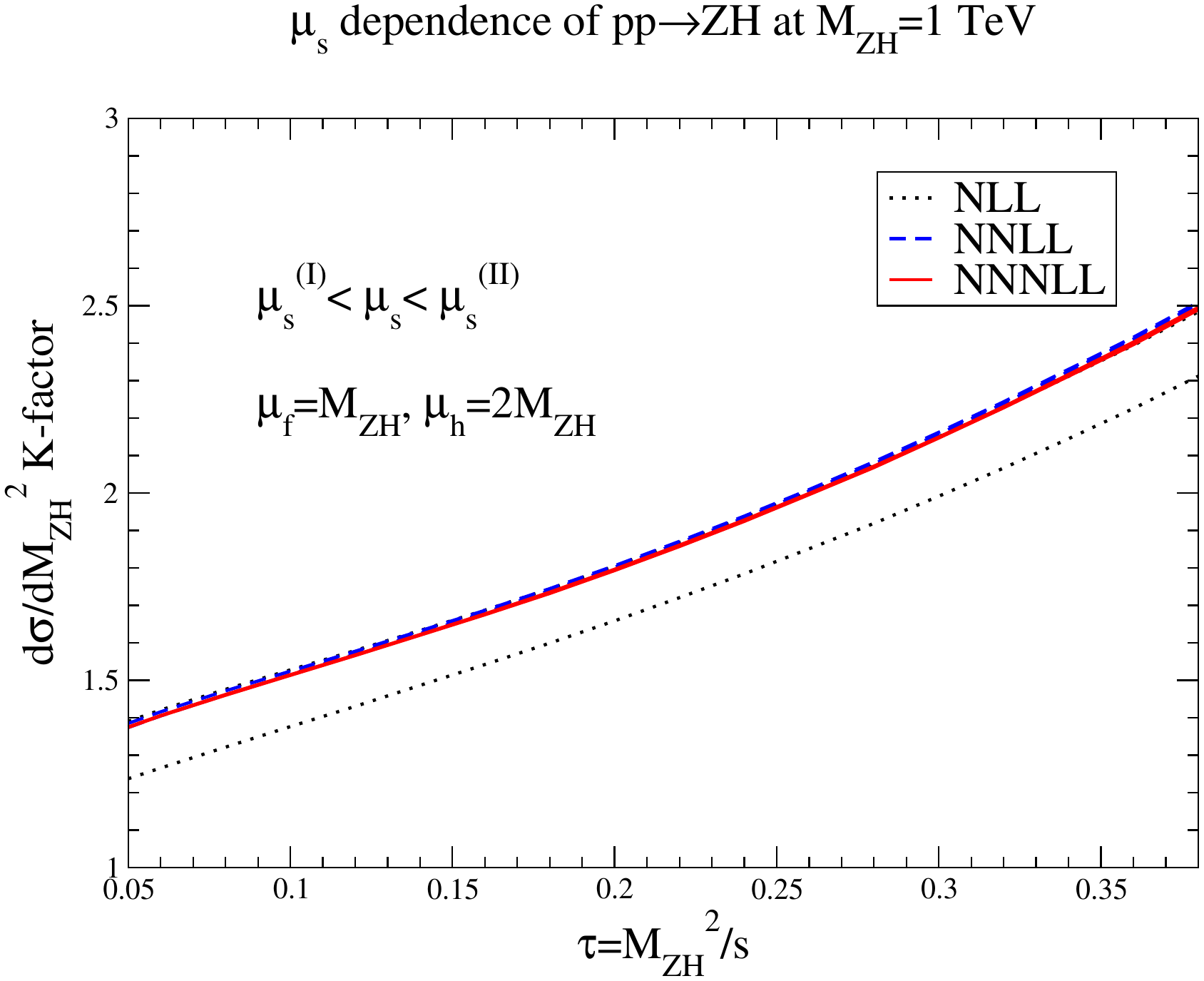}
\label{fg:zhmus}}
\subfigure[]{
\includegraphics[width=0.48\textwidth,clip]{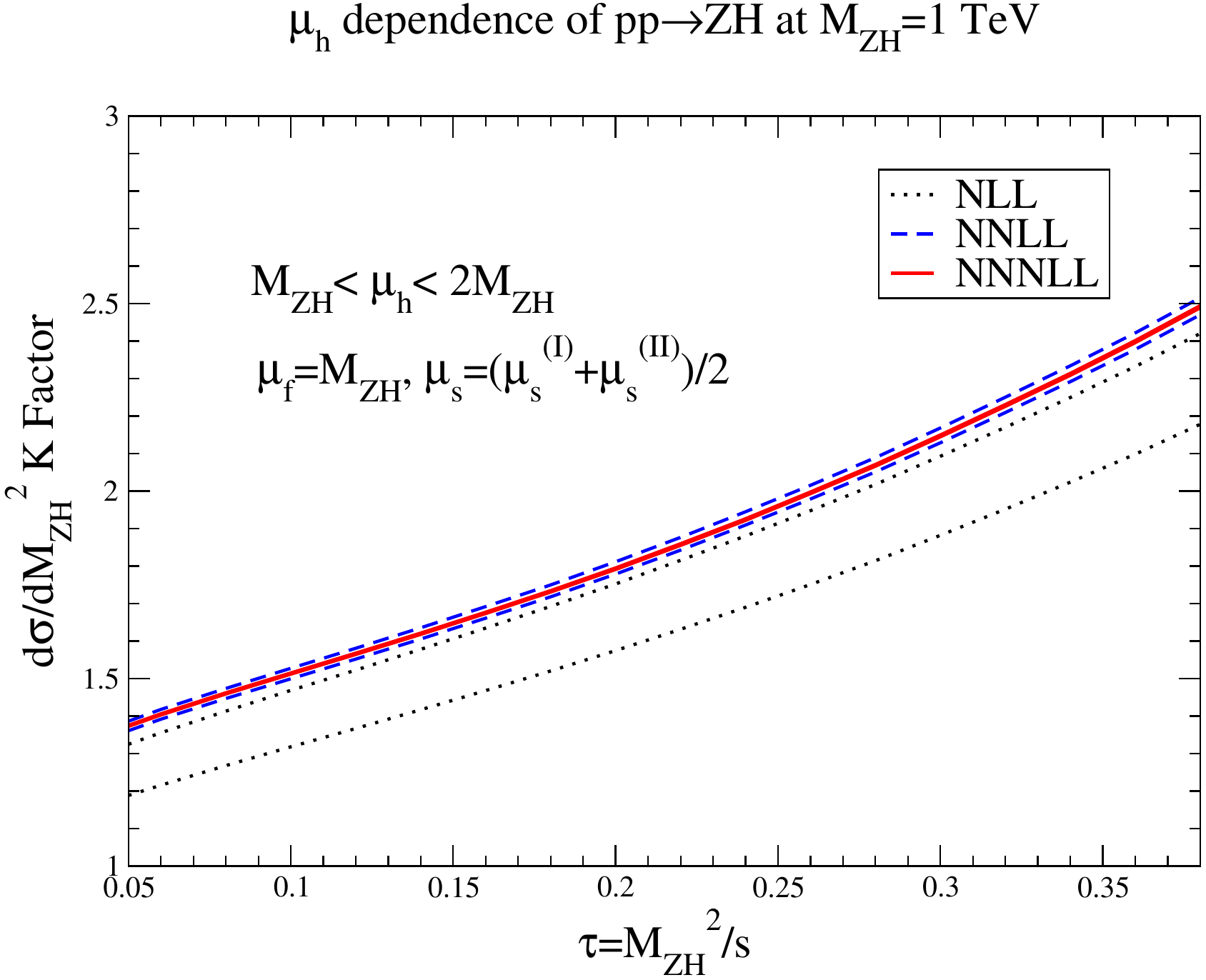}
\label{fg:zhmuh}}
\subfigure[]{
\includegraphics[width=0.52\textwidth,clip]{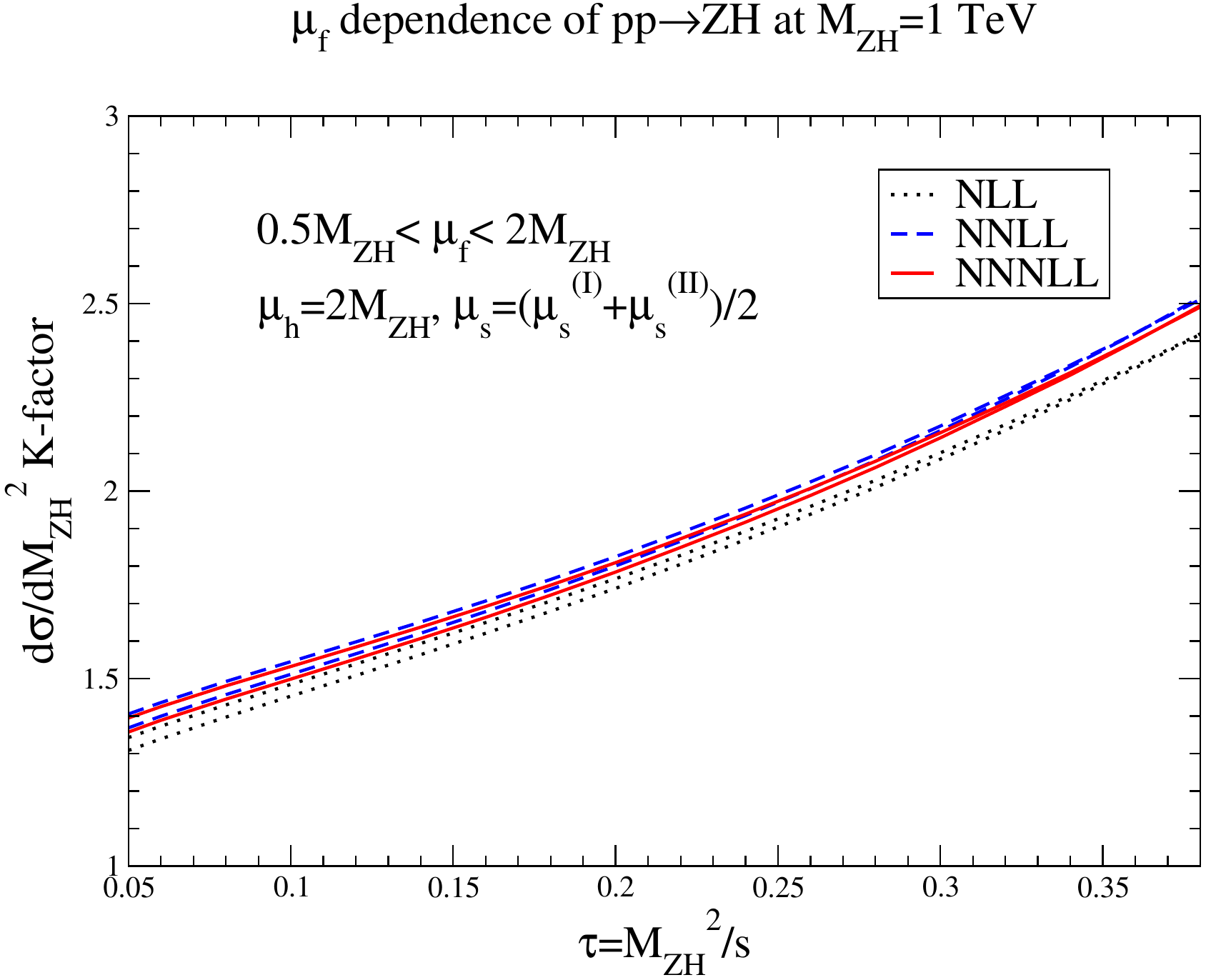}
\label{fg:zhmuf}}

\caption[]{The (a) soft scale, (b) hard scale, and (c) factorization scale dependence of the threshold resummed cross section for
$pp\rightarrow ZH$ at NLL between the dotted lines dotted, NNLL between the dashed lines, and NNNLL between the solid lines normalized to the LO result (the $K$-factor is defined in Eqs.~(\ref{eq:K}) and (\ref{eq:Kthresh})).  The invariant mass $M_{ZH}$ is fixed at $1$ TeV.}
\label{fg:scale}
\end{center}
\end{figure}

To study the scale variation arising from threshold resummation, we investigate the $K$-factor of Eq.~(\ref{eq:K}) defined with 
\begin{equation}
\frac{d\sigma}{dM_{VH}^2}\equiv \biggl[\frac{d\sigma}{dM_{VH}^2}\biggr]_{threshold-resum}.\label{eq:Kthresh}
\end{equation}
To isolate the effects of the scale variation due to threshold resummation from effects of the scale variation due to parton distribution functions (PDFs) and running $\alpha_s$, the $K$-factor is evaluated by using the NNLO MSTW20008~\cite{Martin:2009iq} PDF set and the $3$-loop value of $\alpha_s$ for all orders of the threshold resummed cross section and the LO cross section.
Figure~\ref{fg:scale} shows the scale variation of this choice of $K$-factor as a function of $\tau$  at NLL between the dotted curves, NNLL between the dashed curves, and NNNLL between the solid curves for $ZH$ production at $M_{ZH}=1$~TeV.  
The soft scale variation in $pp\rightarrow ZH$, with $\mu_h$ and $\mu_f$ held constant, is shown in 
Fig.~\ref{fg:zhmus}.  The variation in the NLL result is significant, but
the NNLL and NNNLL curves have little dependence on the soft scale,
justifying the ${\it {ad~ hoc}}$ choices of $\mu_s^{(I,II)}$.  
The $K$-factor grows rapidly as $\tau$ increases, as expected.  The sensitivity
to the hard scale is shown in Fig.~\ref{fg:zhmuh}, with fixed $\mu_s$ and $\mu_f$.  The 
hard scale is set by the invariant mass of the $VH$ pair, and again we find that
at NNLL and NNNLL, there is little dependence on $\mu_h$, showing excellent
convergence of the perturbation series.  Finally, we show the factorization scale 
dependence in Fig.~\ref{fg:zhmuf}. The factorization scale dependence is small even
at NLL. 

We have also considered  the scale dependence of the  matched result for the total cross section. 
Analytic expressions for the LO and NLO
fixed order results are found in Refs.~\cite{Baer:1992vx,Ohnemus:1992bd,Han:1991ia,Brein:2003wg},
 and we use the computer code VH@NNLO for the fixed order NNLO
results.  The matched curves are found using the threshold resummation results of Eq.~(\ref{threshold_combine}). In Fig.~2, we
 use the MSTW2008 $68\%$ confidence
level PDFs, and use LO PDFs for the LO and the NLL-LO matched curves, 
 NLO PDFs for the NLO and NNLL-NLO matched
curves, 
and  NNLO PDFs for the NNLO and NNNLL-NNLO matched curves, and 
use $1,2$ and $3$-loop evolution of $\alpha_s$
respectively.  We include the small contribution from the $gg$ initial state in the $ZH$ NNLO and NNNLL-NNLO matched curves.
  
\begin{figure}
\begin{center}
\subfigure[]{
\includegraphics[width=0.45\textwidth,clip]{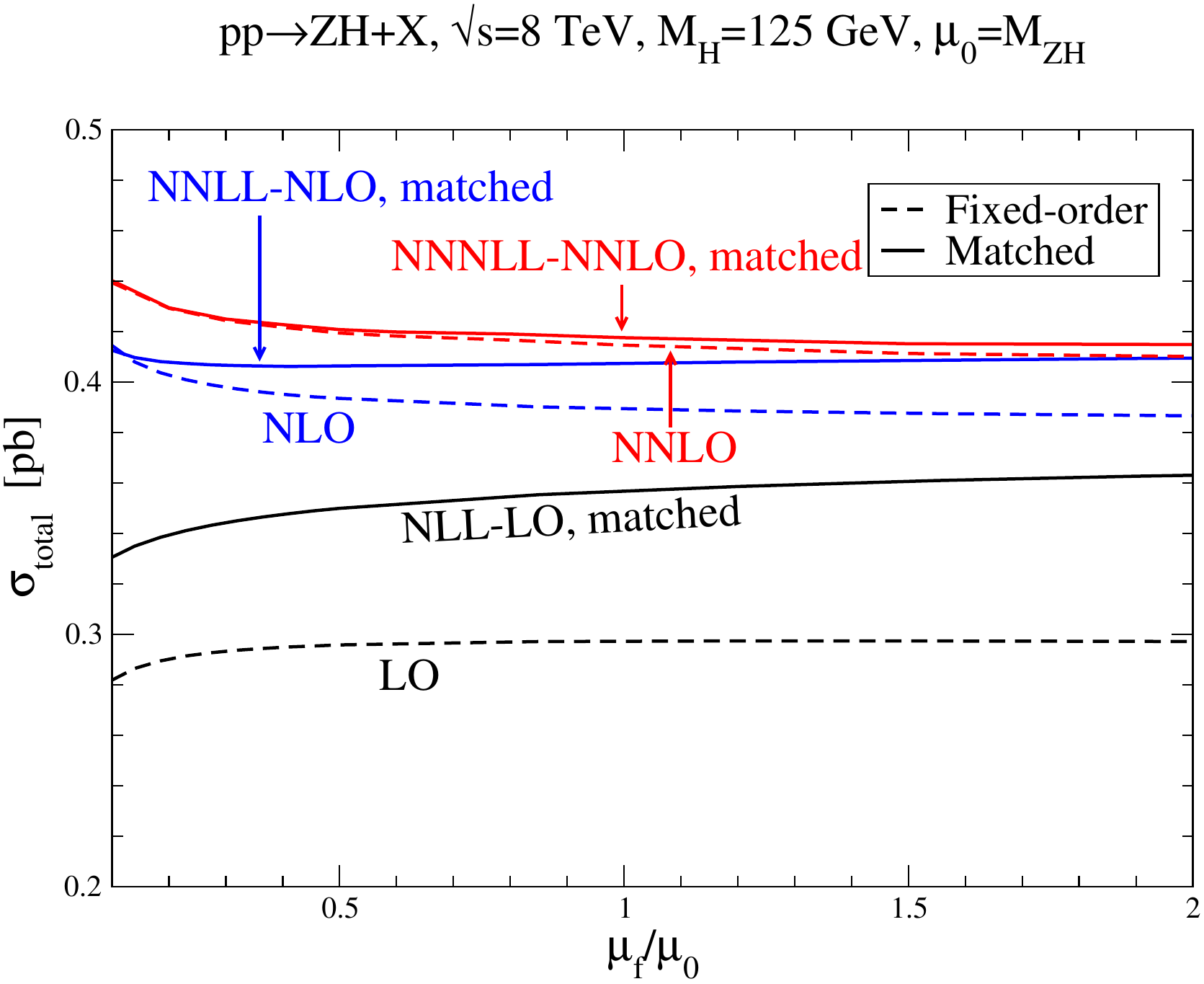}
\label{fg:zh1}}
\subfigure[]{
\includegraphics[width=0.45\textwidth,clip]{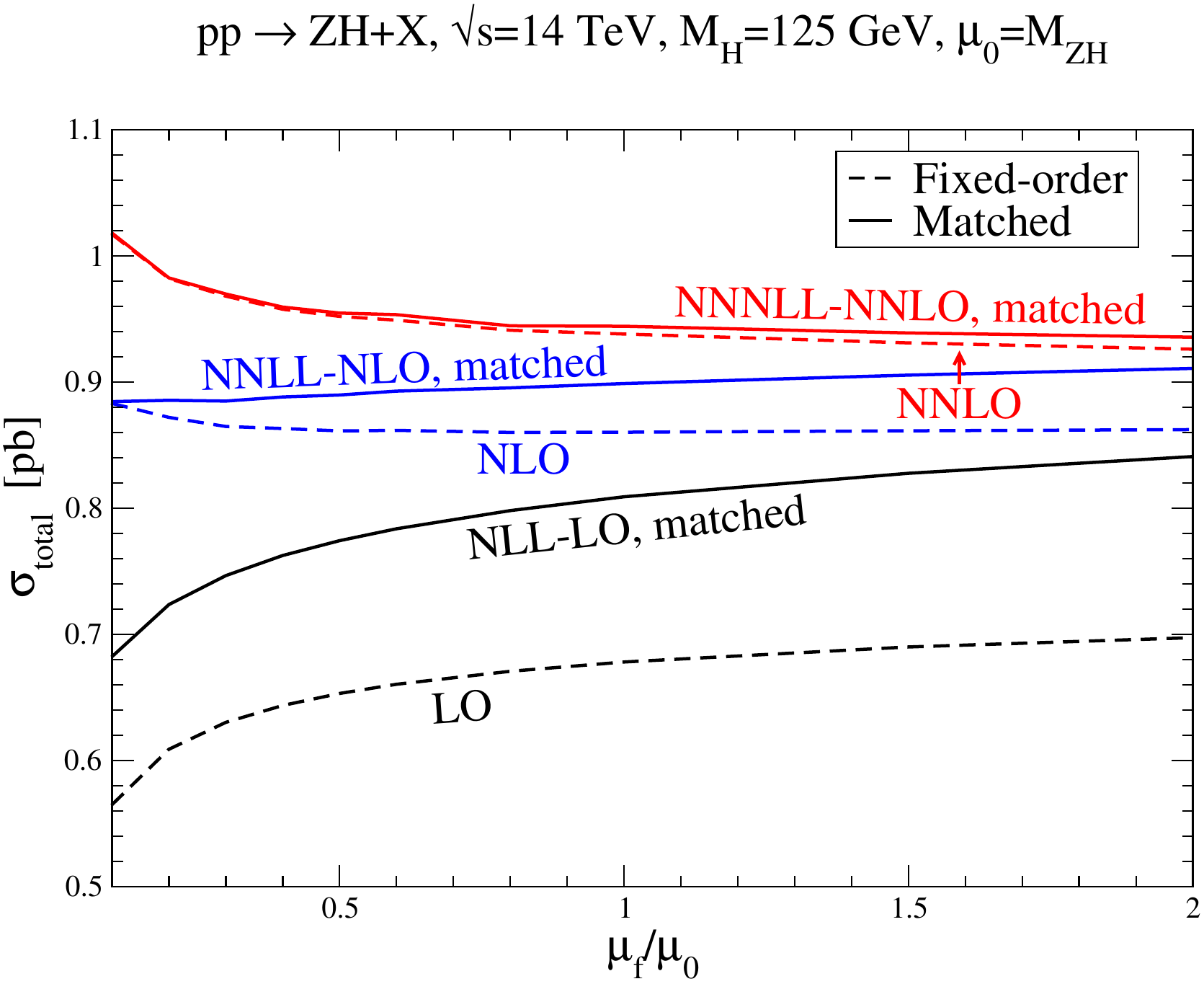}
\label{fg:zh2}}\\
\subfigure[]{
\includegraphics[width=0.45\textwidth,clip]{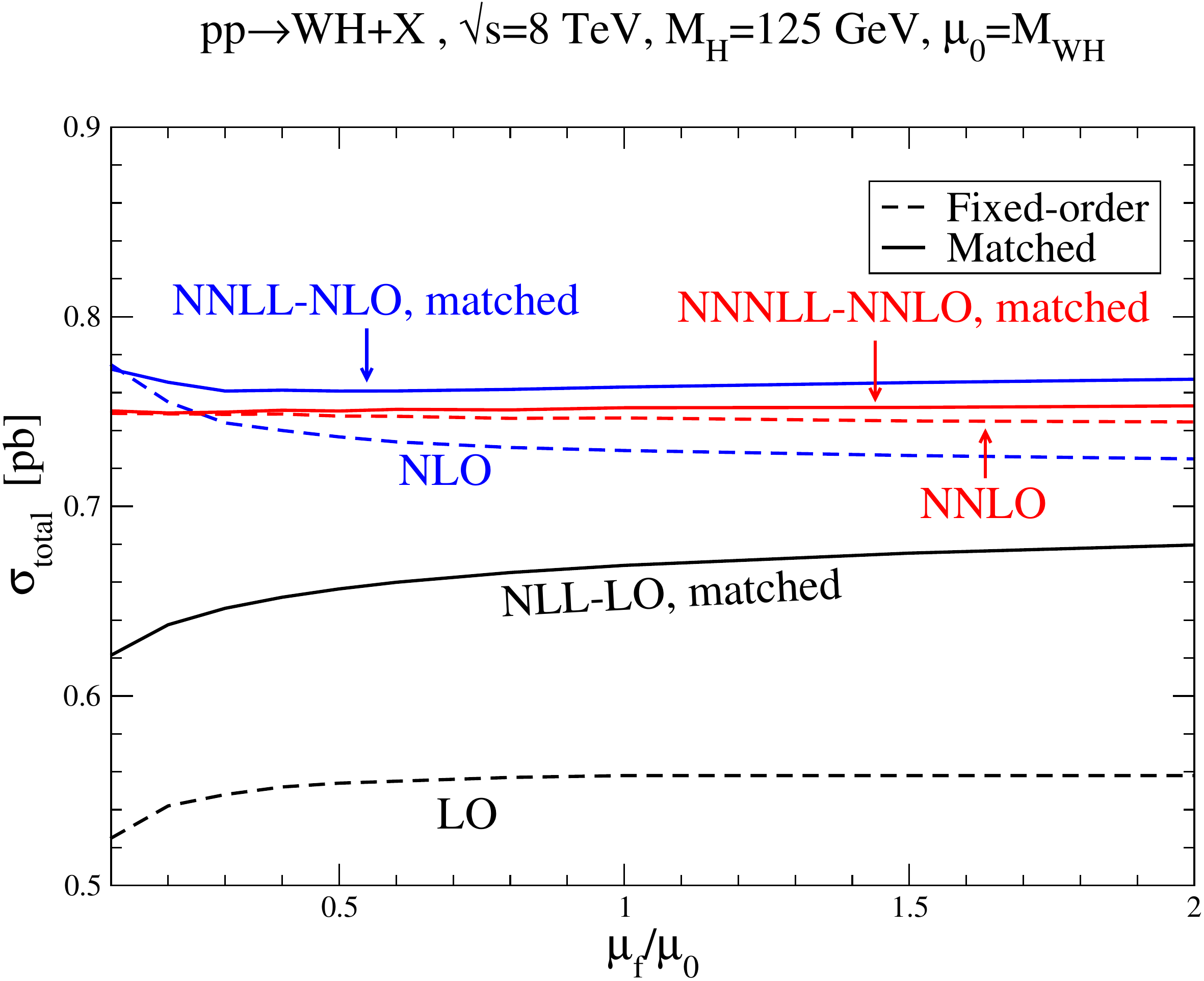}
\label{fg:wh1}}
\subfigure[]{
\includegraphics[width=0.45\textwidth,clip]{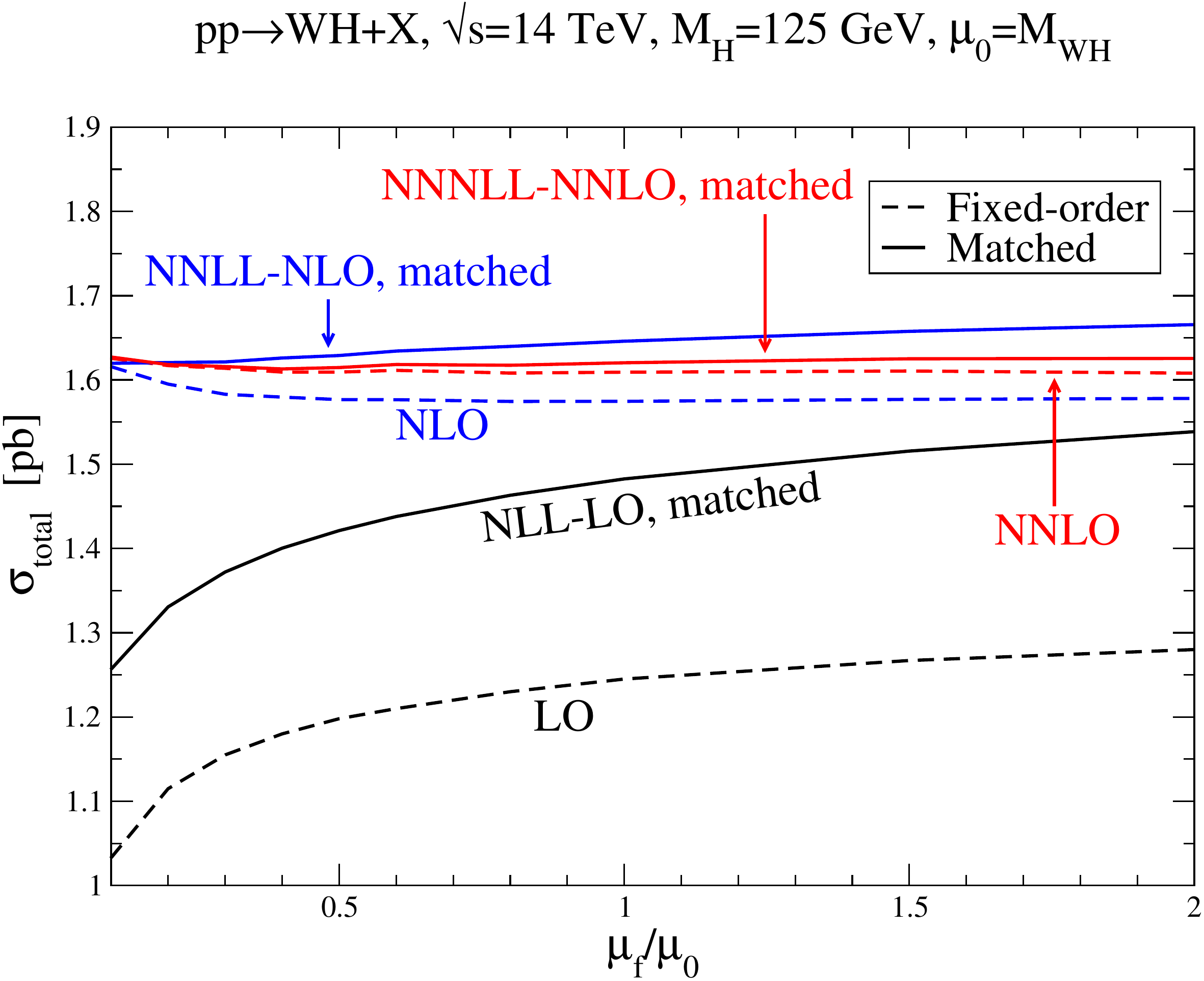}
\label{fg:wh2}}
\caption[]{ Scale dependence of the fixed order (dashed) and threshold resummed matched (solid) cross sections
for (a,b) $ZH$ and (c,d) $WH$ production at  (a,c)$\sqrt{s}= 8$ TeV and (b,d) $\sqrt{s}=14$ TeV.  
The NNLO and NNNLL-NNLO matched $ZH$ results include the contribution from the $gg$ initial state.
 }
\label{fg:vh}
\end{center}
\end{figure}

The results for $ZH$ production at  $\sqrt{s}=8$ TeV and 
$\sqrt{s}=14$ TeV are shown in Figs.~\ref{fg:zh1} and \ref{fg:zh2}, 
respectively. We have chosen
the central scale to be  $\mu_0=M_{ZH}$.
The top and bottom quark loops from the
$gg$ initial state contribute $\sigma_{gg}^{t,b~loops}=0.06$ pb 
at $\sqrt{s}=14$ TeV with $\mu_f=\mu_0$.
This is the reason for the larger splitting between the NLO and NNLO curves
than is seen in the $WH$ results below. 
The fixed-order and matched curves have the renormalization/factorization
scales set equal, $\mu_r=\mu_f$.  The matched and resummed curves have the hard scale, $\mu_h=2M_{VH}$, and 
the soft scale, $\mu_s={1\over 2}(\mu_2^{(I)}+\mu_s^{(II)})$.   The NNNLL-NNLO matched curve is almost identical to the NNLO fixed
order curve, and the resummation has little effect at this order.  On the other hand, the NNLL-NLO matched curve increases the fixed order NLO result (at $\mu_f=\mu_0$) by about $7\%$.

The matched cross sections for $WH$ production at 
$\sqrt{s}=8$ TeV and 14 TeV are shown in Figs.~\ref{fg:vh}(c) and \ref{fg:vh}(d).  These figures show the sum of $W^+H$ and $W^-H$ production.  
As in the $ZH$ case, the NNLO and NNNLL-NNLO matched results
for $WH$ production are quite close and show little scale variation.   The NNLL resummation increases
the NLO fixed order result by $\sim 3\%$.

The uncertainties in the $ZH$ and $WH$ cross sections from PDFs, renormalization and factorization scale dependence, 
and the determination of $\alpha_s$ have been investigated by the LHC Higgs Cross Section Working  Group
 for the NNLO total cross section \cite{Dittmaier:2011ti}.  They find a total uncertainty
 at $\sqrt{s}=8$~TeV of ${\cal{O}}(4\%)$ for $WH$ and ${\cal {O}}(5\%)$ for $ZH$ production for a $125$~GeV Higgs boson. Our results
show that including the resummation of threshold logarithms to NNNLL accuracy does not induce any further uncertainties.
We note that Ref.~\cite{Dittmaier:2011ti} also includes the NLO electroweak effects \cite{Ciccolini:2003jy}, assuming
complete factorization of the QCD and electroweak corrections.  In the $G_\mu$ renormalization scheme, these corrections reduce the total Higgs and vector boson associated rates by about ${\cal{O}}(5\%)$.

\section{Kinematic Distributions}

\subsection{Transverse-Momentum Distributions}
\label{ptdts}

We now give numerical results for the resummed transverse-momentum distributions. 
The distributions are computed at NLL-NLO accuracy with NLO MSTW2008 $68\%
$ confidence level PDFs~\cite{Martin:2009iq} and the $2-$loop evolution of $\alpha_s$
using the formulae of Appendix A.  The numerical results were found by modifying the program HqT2.0~\cite{Bozzi:2005wk,deFlorian:2011xf,HqT}.
 The factorization and renormalization scales 
are set to  the central values of $\mu_f=\mu_r=M_V+M_H$.  
Also, the resummation scale is set equal to the invariant mass of the vector boson 
and Higgs pair, i.e., $Q=M_{VH}$.  

\begin{figure}[tb]
\subfigure[]{
      \includegraphics[width=0.36\textwidth,angle=-90,clip]{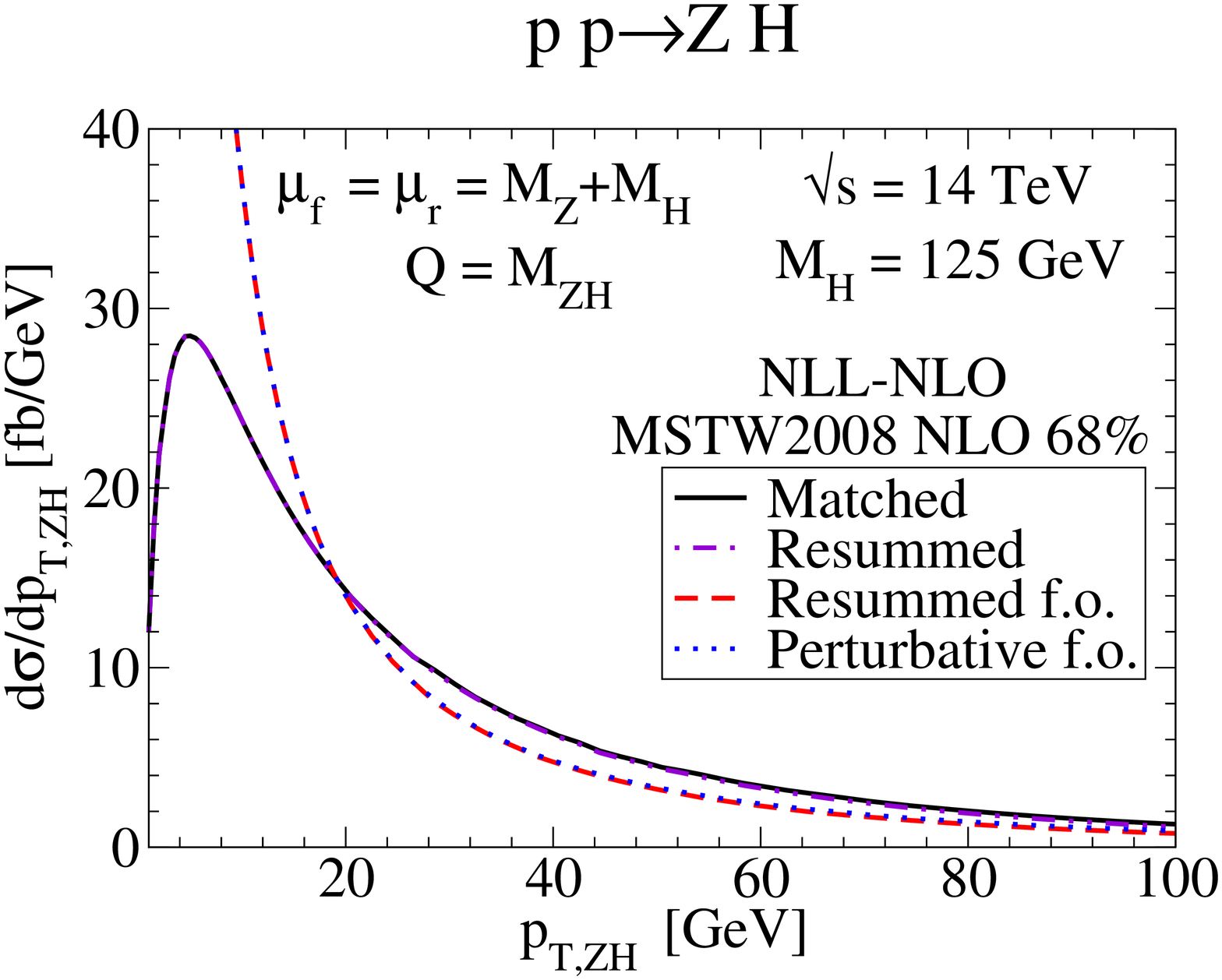}
}
\subfigure[]{
      \includegraphics[width=0.36\textwidth,angle=-90,clip]{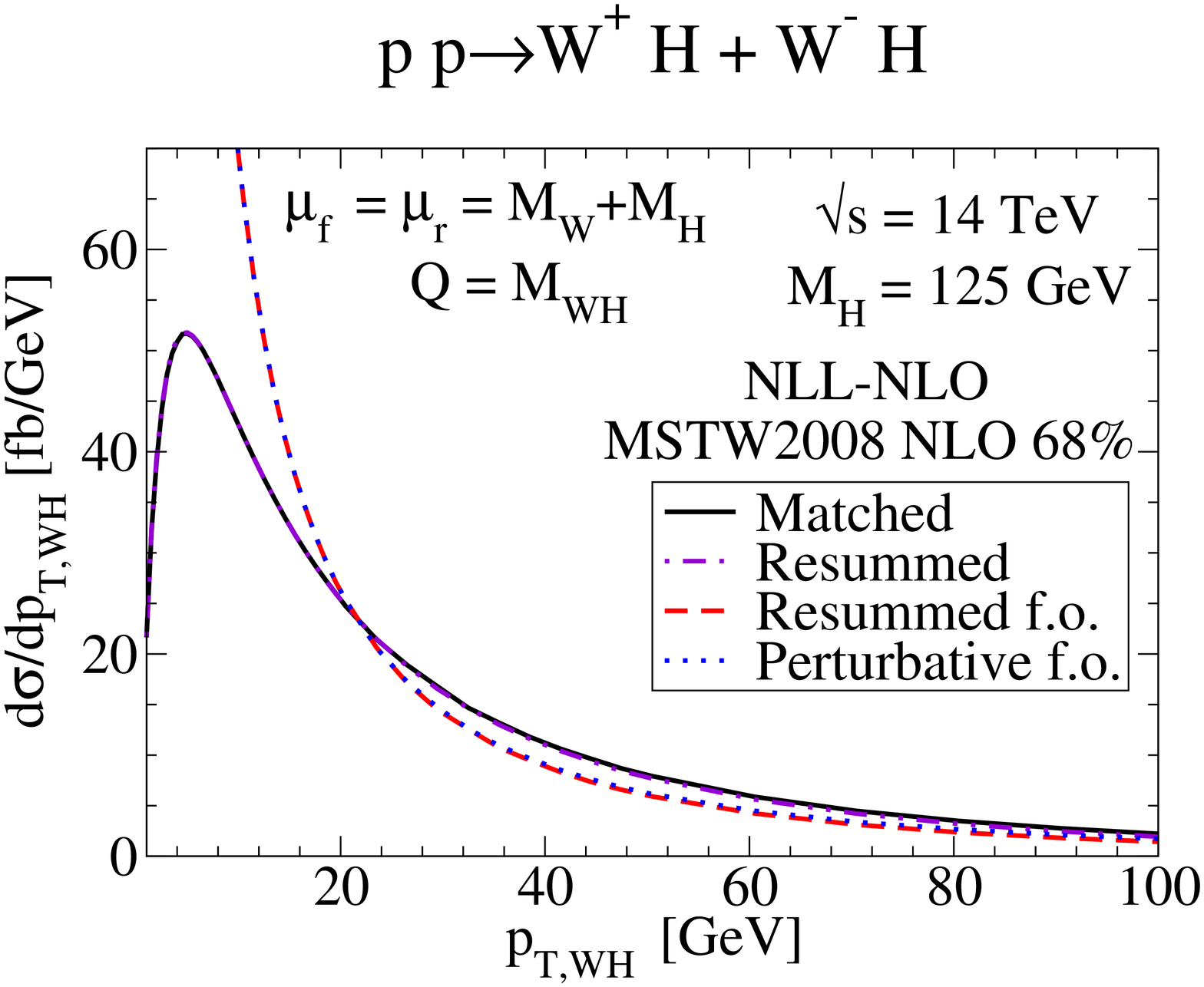}
}
\subfigure[]{\label{ptnorm_ZH.fig}
      \includegraphics[width=0.36\textwidth,angle=-90,clip]{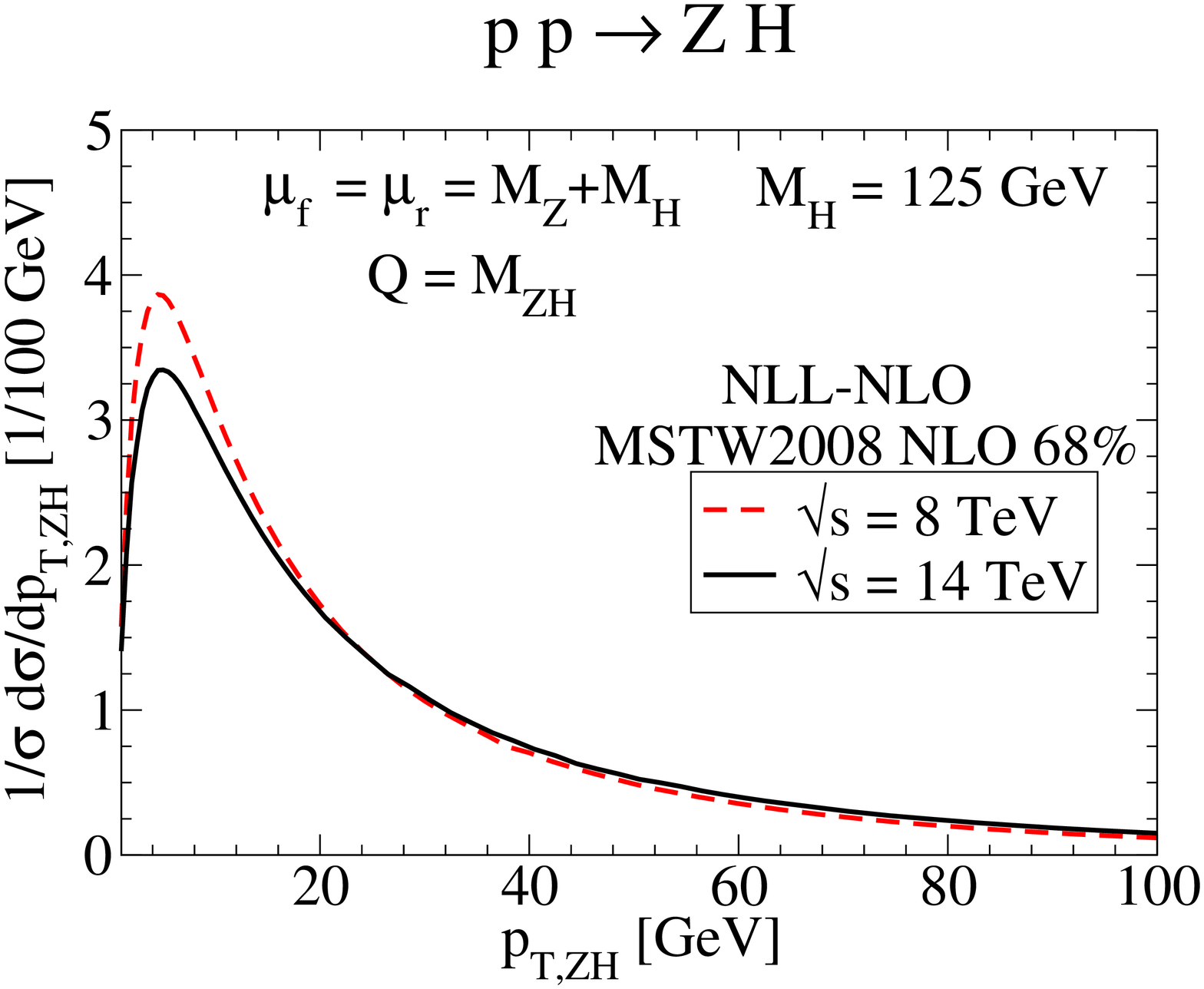}
}
\subfigure[]{\label{ptnorm_WH.fig}
      \includegraphics[width=0.36\textwidth,angle=-90,clip]{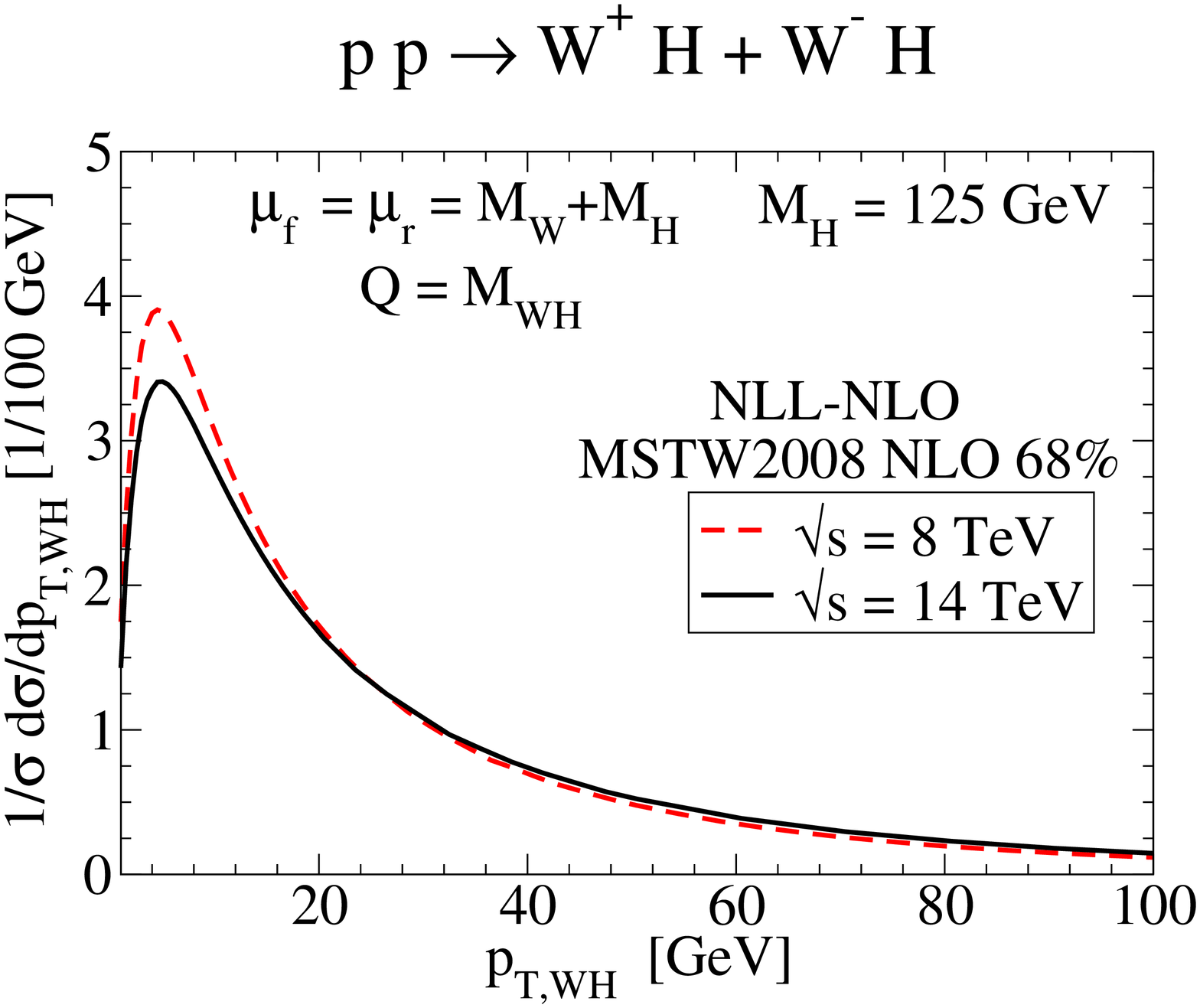}
}
\caption{Transverse-momentum distributions for (a,c) $ZH$, and (b,d) $WH$ 
production at the LHC.  In (a) and (b), the matched distribution is shown
 with a solid line, the resummed distribution with a dot-dash line, the fixed-order expansion of the resummed distribution with a dashed line and the fixed-order perturbative 
 distribution with a dotted line at  $\sqrt{s}=14$ TeV. In (c) and (d), the normalized matched transverse-momentum distributions are shown for both   $\sqrt{s}=8$ TeV 
 (dashed) and $\sqrt{s}=14$ TeV (solid) LHC.}
\label{ptdists.fig}
\end{figure}

Figures \ref{ptdists.fig}(a) and \ref{ptdists.fig}(b) show the transverse-momentum distribution for 
$ZH$ and $WH$ production, respectively, at 
$\sqrt{s}=14$ TeV. The matched transverse-momentum distribution defined by Eqs.~(\ref{trans_1}) and (\ref{finite.EQ}) (solid), resummed (dot-dash), fixed order expansion of the resummed (dashed), and 
fixed-order perturbative (dotted) distributions are shown separately. As expected, the fixed-order expansion of the resummed 
and perturbative distributions are in good agreement.  Hence, the 
finite piece, defined to be the difference between the perturbative distribution and 
fixed-order expansion of the resummed distribution as in Eq.~(\ref{finite.EQ}), is negligible at low transverse momentum and the 
matched distribution is dominated by the resummed contribution.  The transverse-momentum 
distribution is peaked around $5$ GeV for both $WH$ and $ZH$ production.

For comparison, in Figs.~\ref{ptdists.fig}(c) and \ref{ptdists.fig}(d) we present the normalized matched
transverse-momentum distributions for $ZH$ and $WH$ production, respectively, 
at both $\sqrt{s}=8$ TeV (dashed) and 
$\sqrt{s}=14$ TeV (solid).  The position of the peak of the 
transverse distribution is not significantly different between the two LHC energies.  
However, the distribution at  $\sqrt{s}=14$ TeV has a longer tail than at $\sqrt{s}=8$ TeV.
This can be understood by noting that higher transverse-momentum 
events correspond to higher partonic center-of-mass energies.
  Since events with higher partonic center-of-mass energies are 
more easily accessible at $\sqrt{s}=14$ TeV 
 than at $\sqrt{s}=8$ TeV, we would expect there to be larger fraction of high $p_{T,VH}$ 
events at $\sqrt{s}=14$ TeV than at $\sqrt{s}=8$ TeV.  Hence, the transverse-momentum
distribution has a longer tail for $\sqrt{s}=14$~TeV.

\begin{figure}[tb]
\subfigure[]{\label{ptint_ZH.fig}
      \includegraphics[width=0.36\textwidth,angle=-90,clip]{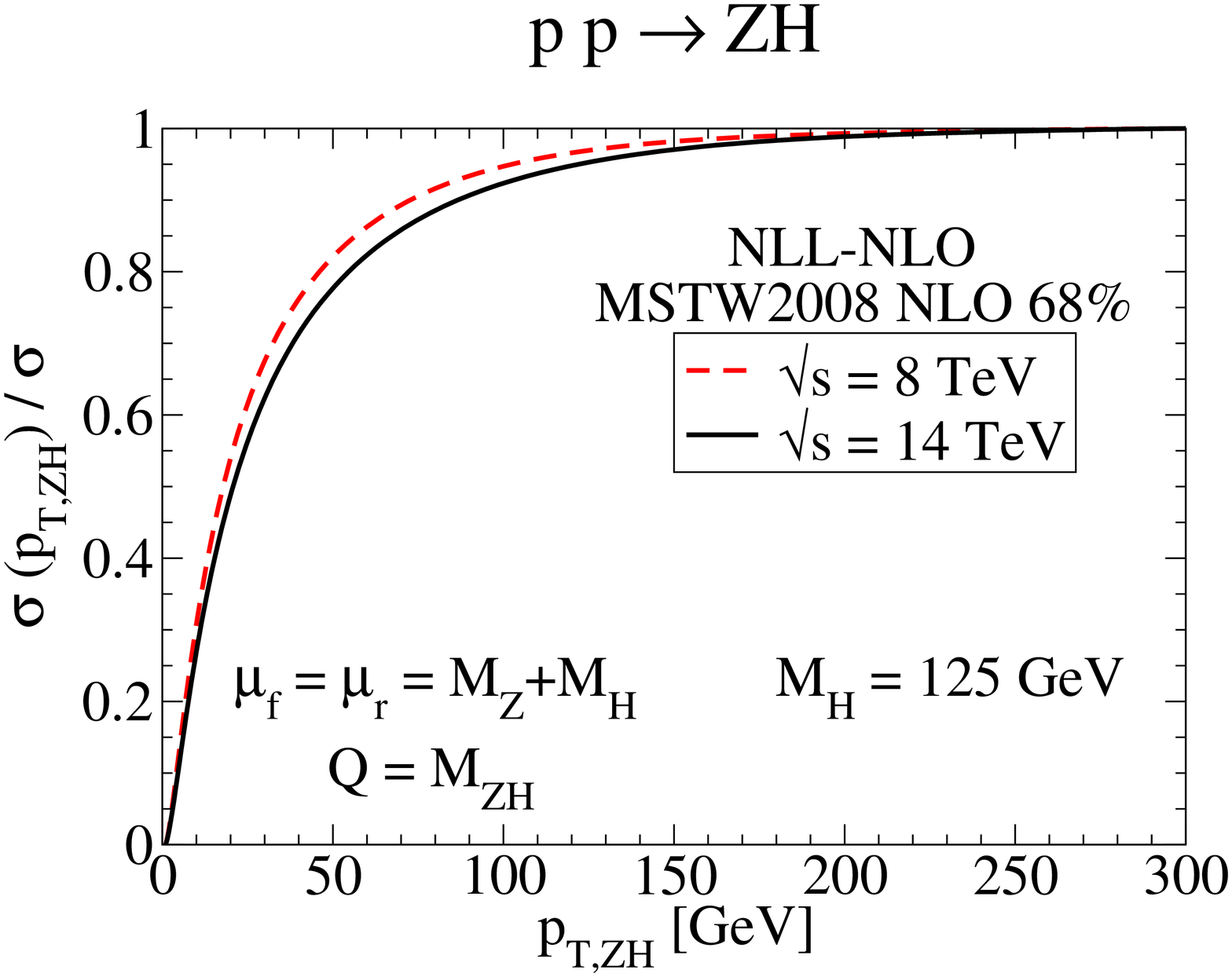}
}
\subfigure[]{\label{ptint_WH.fig}
      \includegraphics[width=0.36\textwidth,angle=-90,clip]{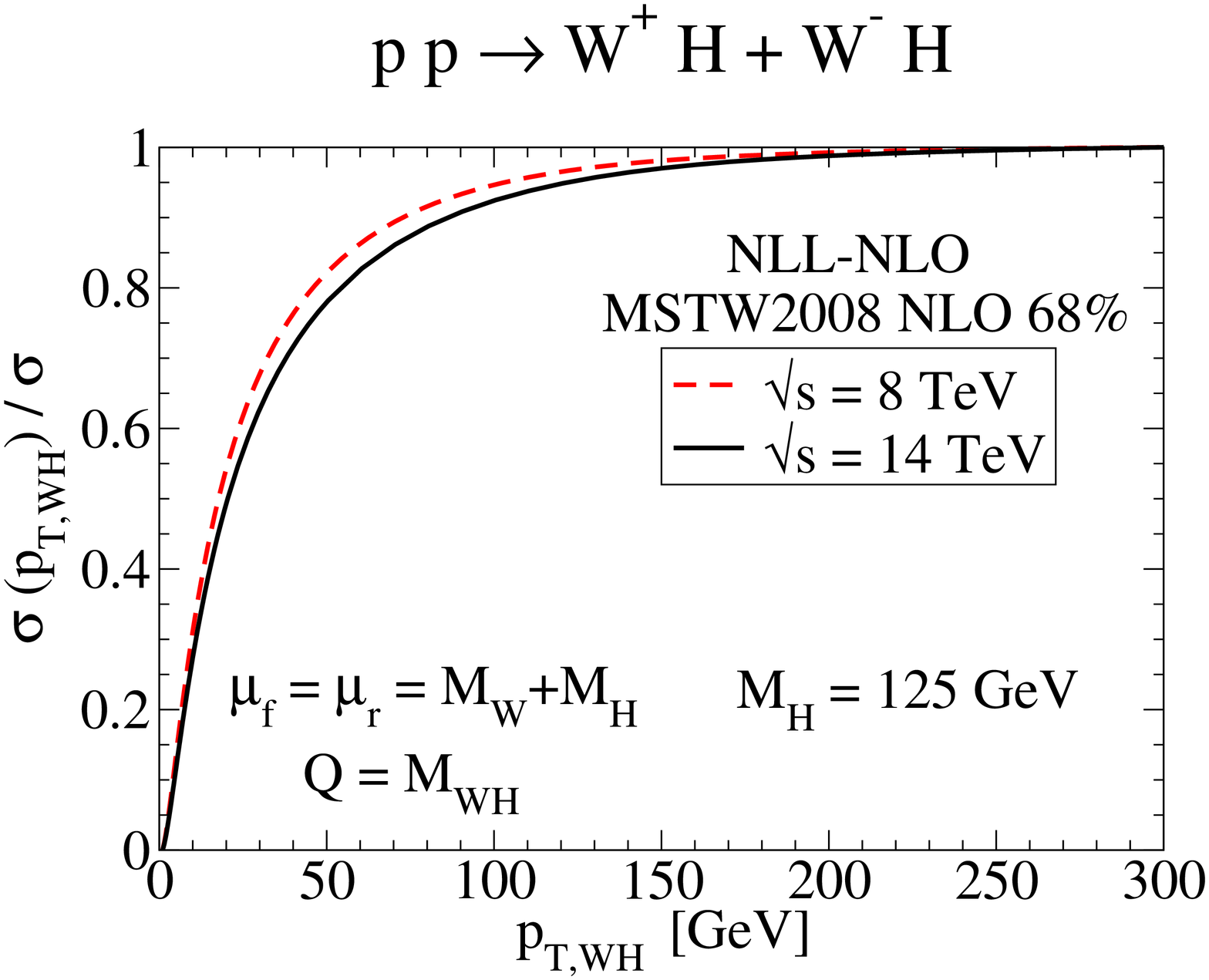}
}
\caption{Integrated matched transverse-momentum distributions normalized to the total 
cross section for both (a) $ZH$ and (b) $WH$ production.  Results 
for  $\sqrt{s}=8$ TeV and $\sqrt{s}=14$ TeV are shown with dashed and solid lines, respectively.}
\label{ptint.fig}
\end{figure}

Finally, we comment how the transverse-momentum resummation can effect the 
analysis of kinematical cuts on the signal cross section, particularly in relation to jet vetoes.  At hadron machines, the 
$VH$ production with Higgs decaying to $b\bar{b}$ has large QCD backgrounds.  
To reduce the backgrounds and effectively trigger on the signal, one usually considers
 leptonic decays of the vector boson.  However, if the vector boson decay 
contains missing energy, $W\rightarrow\ell\nu$ or 
$Z\rightarrow\nu{\overline{\nu}}$, 
semileptonic decays of $t\bar{t}$ can be a significant background.  Since the 
$t\bar{t}$ background typically has more hard jets than the
 $VH$ signal, a jet veto may be 
applied to suppress this background.  We note that vetoing jets with a minimum 
transverse momentum can be approximated by placing an upper limit on the $VH$
 transverse momentum and, as can be seen in Figs.~\ref{ptdists.fig}(a) and \ref{ptdists.fig}(b), 
 the perturbative calculation is unreliable in this regime.  Hence, to fully account
for the effects of a jet veto the soft-gluon resummation is needed.  There has
 been much recent work on the systematic resummation of the large 
logarithms  associated with jet vetoes~\cite{Berger:2010xi,Banfi:2012yh,Banfi:2012jm,Becher:2012qa,Tackmann:2012bt}.

To approximate the effect  on the total cross section of a veto on jets with 
transverse momentum larger than $p_{T,VH}$ we define
\begin{eqnarray}
\sigma(p_{T,VH}) = \int^{p_{T,VH}}_0 dq_{T,VH }\frac{d\sigma}{dq_{T,VH}},
\label{xsect_veto.EQ}
\end{eqnarray}
where $d\sigma/dq_{T,VH}$ is the matched transverse-momentum 
distribution at NLL-NLO in Eqs.~(\ref{trans_1}) and (\ref{finite.EQ}).  
Figure~\ref{ptint.fig} shows this cross section normalized to the total $p_{T,VH}$ resummed and matched 
cross section as a function of $p_{T,VH}$ for (a) $ZH$ and (b) $WH$ production for 
both $\sqrt{s}=8$ TeV (dashed) and $\sqrt{s}=14$ TeV (solid).  As noted before in the 
discussion of Figs.~\ref{ptdists.fig}(c) and \ref{ptdists.fig}(d), 
at $\sqrt{s}=14$ TeV, there is expected to be a 
larger fraction of high transverse-momentum jets 
than at $\sqrt{s}=8$ TeV.  Hence, $\sigma(p_{T,VH})/\sigma$ grows 
more slowly at $\sqrt{s}=14$ TeV than at $\sqrt{s}=8$ TeV.  From the figures we
 see that the effects of a $20$ ($30$) GeV $p_{T,VH}$ cut decreases the NLO cross 
 section by $\sim45\%$ ($\sim33\%$) and $\sim50\%$ ($\sim37\%$) 
at  $\sqrt{s}=8$ TeV and  $\sqrt{s}=14$ TeV, respectively.

\subsection{Invariant-Mass Distributions}
\label{dts}

\begin{figure}[tb]
\subfigure[]{
      \includegraphics[width=0.45\textwidth,angle=0,clip]{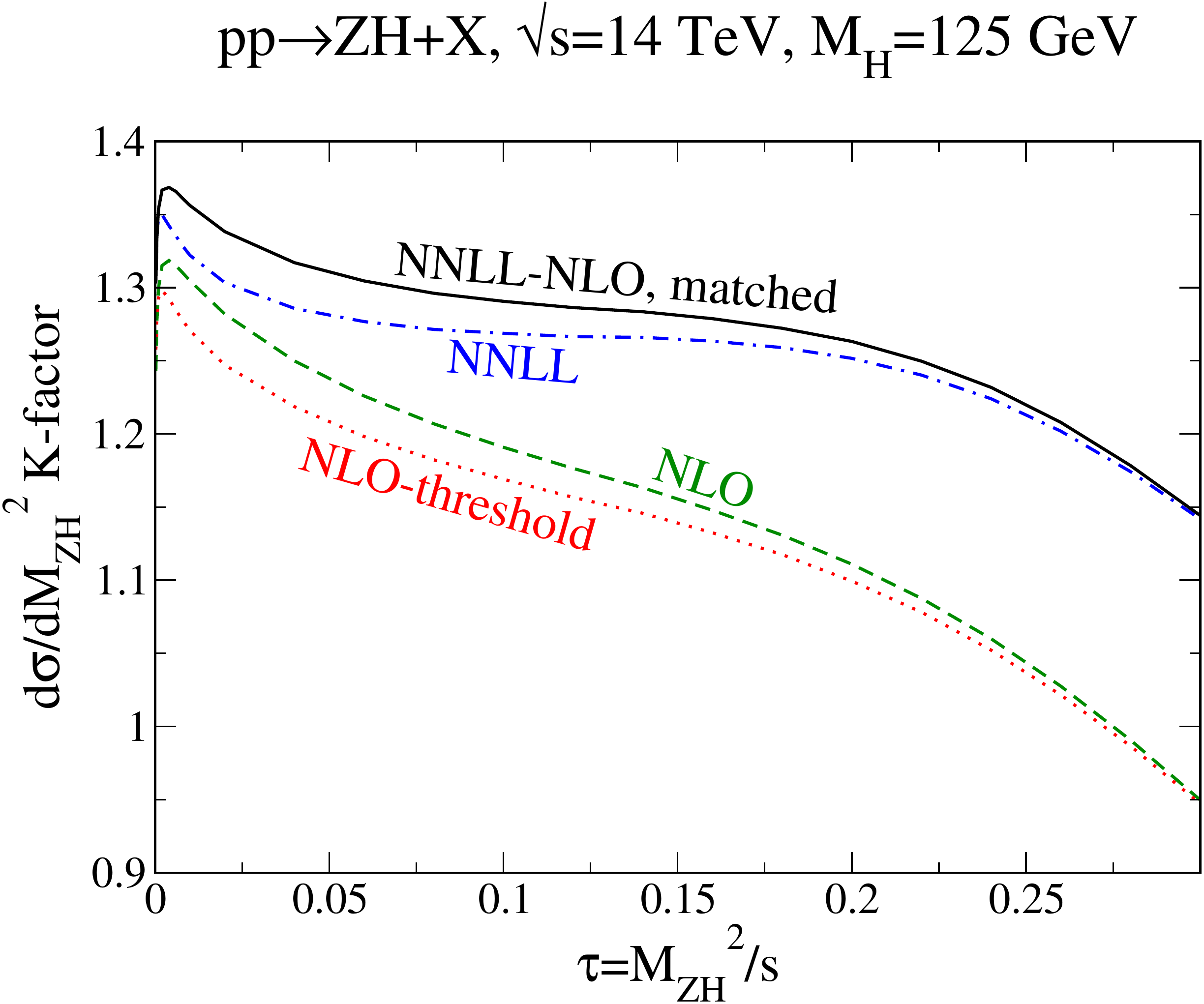}
}
\subfigure[]{
      \includegraphics[width=0.45\textwidth,angle=0,clip]{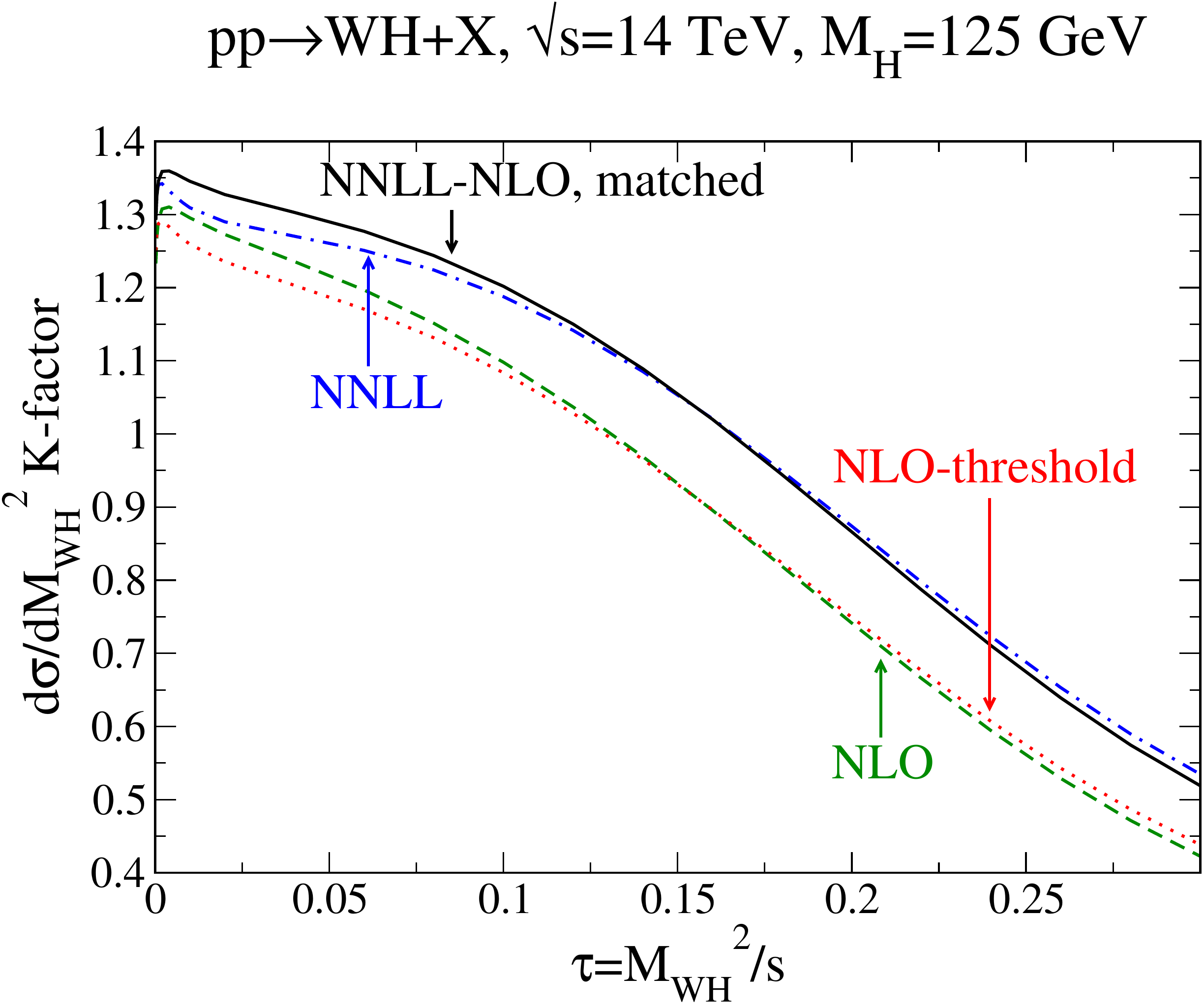}
}
\caption{$K$-factor distributions at $\sqrt{s}=14$~TeV for (a) $ZH$ and (b) $WH$ production.  The NNLL-NLO matched result is shown with solid lines, the NNLL threshold resummed result with dot-dashed lines, 
the leading threshold singularity of the NLO fixed-order result with 
dotted lines, and the NLO fixed-order result with dashed lines.}
\label{fg:HV_14TeV_M}
\end{figure}

In this section, we give numerical results for the invariant-mass distributions
including threshold resummation and matching, using the analytic
formulae of Appendix B. Since the distributions vary over many orders of magnitude, it is easier to see the effects in the $K$-factor, as defined in Eq.~(\ref{eq:K}). 
Figures \ref{fg:HV_14TeV_M}(a) and \ref{fg:HV_14TeV_M}(b) show the $K$-factor versus $\tau$ at NNLL-NLO with $\sqrt{s}=14~\textrm{TeV}$ for $pp\rightarrow ZH+X$ and $pp\rightarrow WH+X$, respectively.  The $K$-factor for the matched result of Eq.~(\ref{threshold_combine}) is shown with solid lines, the threshold-resummed contribution with dot-dashed lines, the fixed-order perturbative contribution with dashed lines, and the contribution from the leading threshold singularity of the fixed-order perturbative piece with dotted lines. Here we use MSTW2008 $68\%$ confidence level PDFs~\cite{Martin:2009iq}. The scales are chosen to be $\mu_f=M_{ZH}$, $\mu_h=2M_{ZH}$ and $\mu_s=\frac{1}{2}(\mu_s^I+\mu_s^{II})$ as in Section~\ref{sigres}. For the NLO fixed-order result, the leading threshold singularity of the NLO fixed-order result and the threshold-resummed result at NNLL, the NLO PDFs and 2-loop $\alpha_s$ are used, whereas for the LO fixed-order denominator of the $K$-factor, we use the LO PDFs and 1-loop $\alpha_s$. As expected, the leading singularity and fixed-order results (the two lower curves) are close to each other, since the leading singularity dominates in the fixed-order result. On the other hand, the resummation effect is significant at high $\tau$, as seen by the large enhancement of the NNLL (the two upper curves) from the NLO result of $\sim 20\%$ for both $ZH$ and $WH$ at $\tau=0.3$. 

\begin{figure}[tb]
\subfigure[]{
      \includegraphics[width=0.45\textwidth,angle=0,clip]{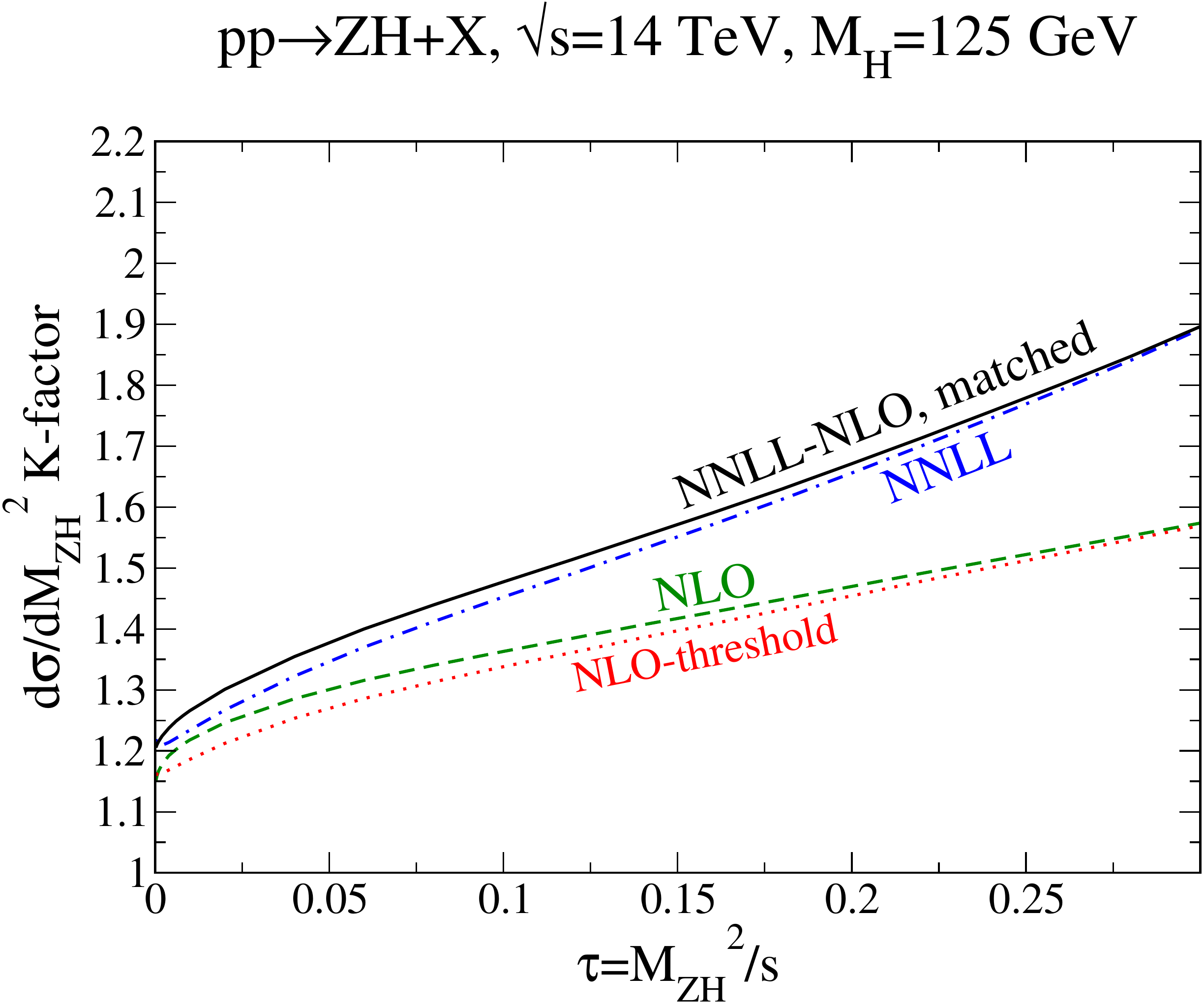}
}
\subfigure[]{
      \includegraphics[width=0.45\textwidth,angle=0,clip]{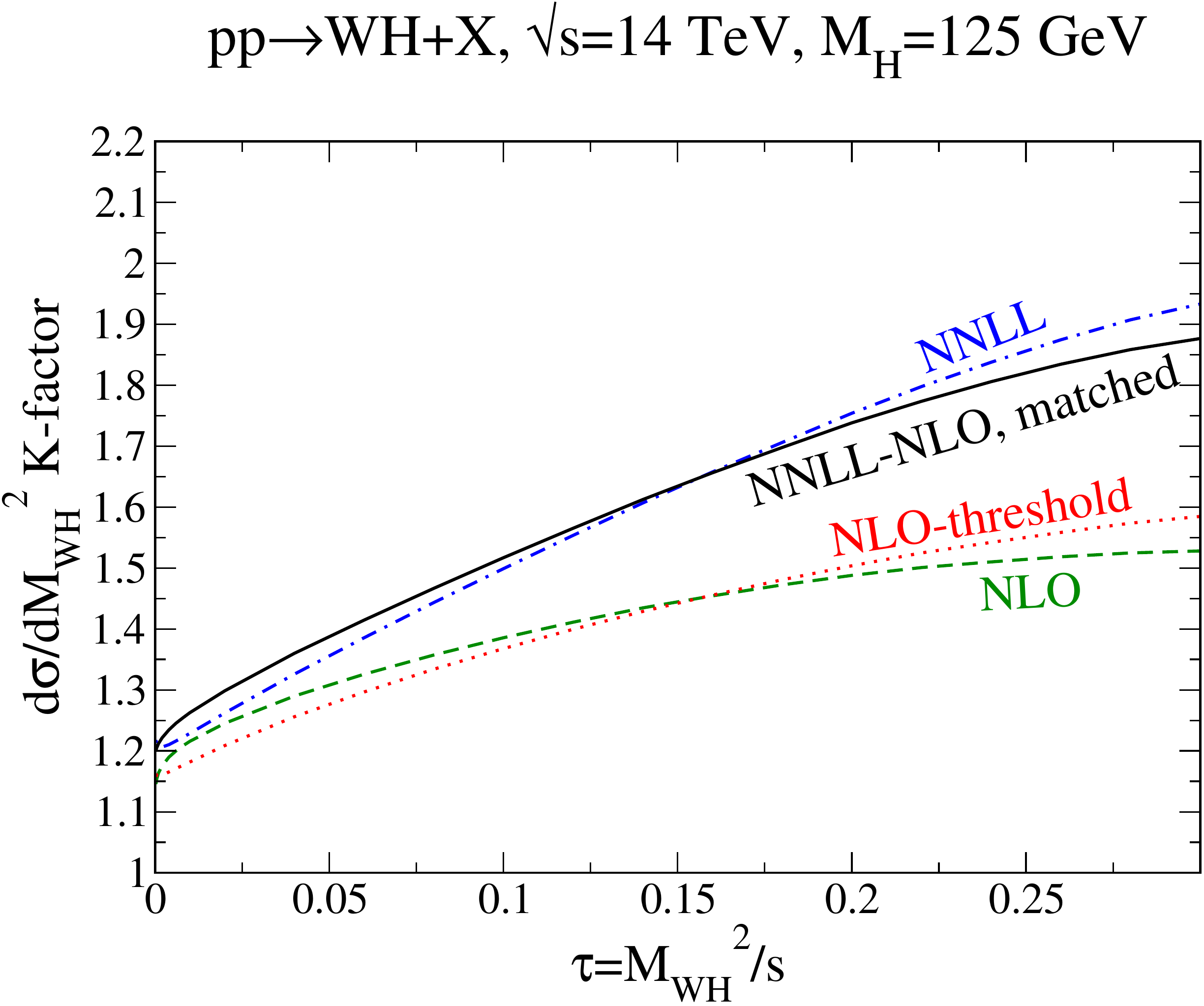}
}
\caption{$K$-factor distributions at $\sqrt{s}=14$~TeV for (a) $ZH$ and (b) $WH$ production.  The NNLL-NLO matched result is shown with solid lines, the NNLL threshold resummed result with dot-dashed lines, the leading threshold singularity of the NLO fixed-order result with dashed lines, and the NLO fixed-order result with dotted lines. The NLO PDFs and 2-loop $\alpha_s$ are adopted for all the results as well as the LO denominator.}
\label{fg:HV_14TeV_M_nlopdf}
\end{figure}

The decrease of the $K$-factor at higher $\tau$ values is due to the PDF effect. To see this, we artificially adopt the NLO MSTW2008 $68\%$ confidence level PDFs and 2-loop $\alpha_s$ for the NLO fixed-order result, the leading threshold singularity of the NLO-fixed-order result and the threshold-resummed result at NNLL, as well as the LO denominator, and show the $K$-factors of these results with $\sqrt{s}=14~\textrm{TeV}$ for $pp\rightarrow ZH+X$ and $pp\rightarrow WH+X$ in Fig.~\ref{fg:HV_14TeV_M_nlopdf}(a) and~\ref{fg:HV_14TeV_M_nlopdf}(b) respectively. This is to isolate the effects of PDFs from a dynamical origin. The choice of scales is the same as in Fig.~\ref{fg:HV_14TeV_M}. 
We note that the monotonic increase of the $K$-factor distributions in Fig.~\ref{fg:HV_14TeV_M_nlopdf} is drastically different from that in Fig.~\ref{fg:HV_14TeV_M}. This demonstrates the importance of a consistent choice of PDFs as in Fig.~\ref{fg:HV_14TeV_M}.

\begin{figure}
\begin{center}
\includegraphics[scale=0.4]{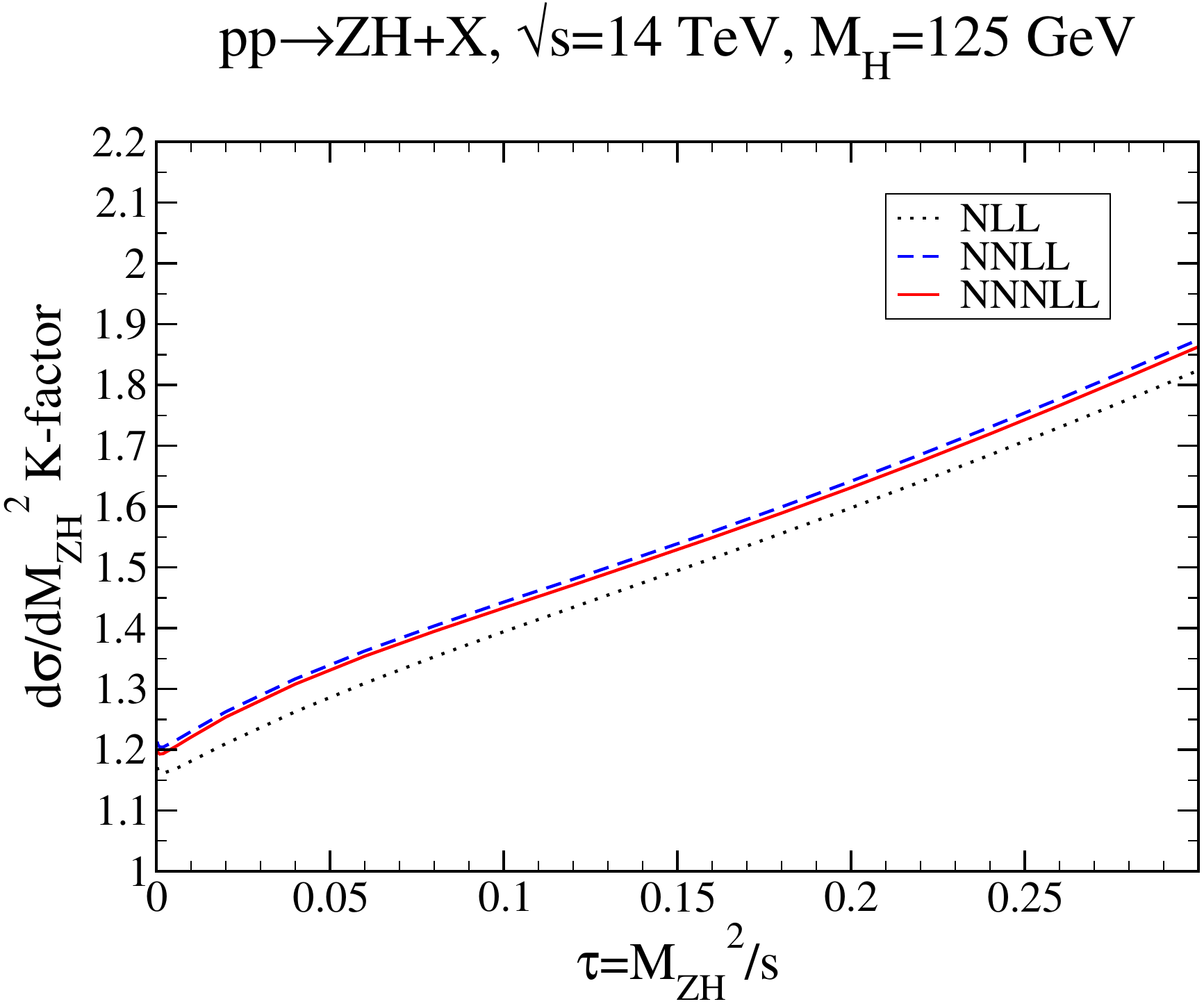}
\vskip .05in  
\caption[]{$pp\rightarrow ZH+X$ $K$ factor distribution at $\sqrt{s}=14~\textrm{TeV}$ for the threshold resummed piece at various orders of the logarithmic approximation, using the same PDFs for all curves.}
\label{fg:HZ_14TeV_M_resum}
\end{center}
\end{figure}

To examine the convergence of the perturbative series, we plot the $K$-factors for the resummed results at NLL, NNLL and NNNLL with $\sqrt{s}=14$ TeV for $pp\rightarrow ZH+X$ in Fig.~\ref{fg:HZ_14TeV_M_resum}, 
using NNLO MSTW2008 $68\%$ confidence level PDFs and 
3-loop $\alpha_s$ for all the resummed results as well as the LO 
denominator. We see from Fig.~\ref{fg:HZ_14TeV_M_resum} that the difference between NNLL and NNNLL is tiny ($<1\%$), confirming the excellent convergence of the perturbative series at this order especially after leaving out the PDF effect.

\section{Conclusions}
\label{conc}

Given the exciting discovery of a Higgs-like particle
 at the LHC~\cite{atsem,cmssem}, it becomes imperative to determine its properties. 
Thus its production rate at the LHC must be calculated as accurately as possible. 
Since the gauge boson--Higgs associated production is one of the channels 
that unambiguously probes the $VVH$ coupling with $V=W^{\pm}$ or $Z$, it is of particular interest.
We  combined the long-known fixed-order perturbative QCD calculations
for $VH$ production \cite{Brein:2003wg} 
with soft-gluon resummation of both threshold logarithms 
and logarithms which are important at low transverse 
momentum of the $VH$ pair.

After a brief overview of the resummation formalism, 
we carried out detailed numerical analyses at the LHC for 
$\sqrt{s}=8$ TeV and 14 TeV. 
The overall corrections from NNLO fixed order calculations are sizable, increaing the LO rate by a factor as large as about $30\%$ \cite{Dittmaier:2011ti}.
After implementing threshold resummation,
the dependence of the cross section and various kinematic distributions 
on the soft and hard scales, 
as well as on the factorization scale is very weak, 
indicating the reliability of the  calculations. 
The NNLL threshold resummed total cross section  increases the 
fixed-order NLO  result by about $7\%$, 
while the NNNLL resummed result has little impact on the NNLO  fixed order rate,
demonstrating the  excellent convergence of the perturbation series. 

The transverse-momentum spectrum of the $VH$ system is 
calculated via soft and collinear gluon resummation. 
The distribution is peaked near 5 GeV and the spectrum is 
slightly harder at the center-of-mass energy of 14 TeV than at 8 TeV. 
Using the matched transverse-momentum distribution, we have also calculated the effect on the NLO cross section of placing an upper bound on the $p_{T}$ of the $VH$ system.  Since such an upper bound on the transverse momentum of the $VH$ system limits the amount of transverse momentum a jet may carry in $VH+X$ events, we expect the upper bound on the 
$p_{T,VH}$ of the $VH$ system to approximate a jet veto.

As a final remark, our calculations can be easily extended to 
other electroweak pair production processes  with the same color structures 
which arise via $q\bar q'$ annihilation at leading order, such as the EW gauge boson pairs and the Higgs pair production 
$H^{0}A^0,\ H^{0}H^{\pm},\ A^{0}H^{\pm}$ and $H^{+}H^{-}$ \cite{Christensen:2012si}.

\section*{Appendix A: $p_{TV}$ Resummation}
In this appendix, we list the functions needed for the $p_{T,VH}$
resummation of Section \ref{sec_trans} \cite{Kodaira:1982az,Kodaira:1981nh}.  All formulae in this appendix can be found in Ref.~\cite{Bozzi:2005wk}, but we include them for the convenience of the reader.  First, the coefficients of the QCD beta function are normalized according to the expansion
\begin{equation}
{d\ln\alpha_s(\mu^2)\over d\ln\mu^2}=\beta(\alpha_s(\mu^2))=-\sum^{\infty}_{n=0}\beta_n\left({\alpha_s(\mu^2)\over4\pi}\right)^{n+1}\, .
\end{equation}
At LL only the function $g_N^1$ is needed
and the Born level contribution arises only from 
$q {\overline q}^\prime$
scattering\cite{Bozzi:2005wk},
\begin{eqnarray}
&& g_N^1(\alpha_sL) = \biggl(
{4A^1_q\over \beta_0}\biggr)
{\lambda+\ln(1-\lambda)
\over\lambda}\nonumber \\
&& \lambda \equiv
\biggl({\beta_0\over 4\pi}\biggr)\alpha_s(\mu_r)L, \quad
\beta_0= \biggl({33-2n_{f}\over 3}\biggr), \quad
A_q^1= {4\over 3}=C_F\, ,
\end{eqnarray}
and $\L=\ln\biggl({Q^2 b^2\over b_0^2}\biggr)$.

At NLL the functions $g_N^2$ and $H_N^{VH(1)}$ are 
needed\cite{Bozzi:2005wk},
\begin{eqnarray}
g_N^2(\alpha_sL,{M_{VH}\over \mu_r}, 
{M_{VH}\over Q})&=&
{4{\overline {B}}_{q,N}^1\over \beta_0}
\ln(1-\lambda)-{16 A_q^2\over\beta_0^2}
\biggl(
{\lambda\over 1-\lambda}
+\ln(1-\lambda)\biggr)
\nonumber\\
&&+
{4 A_q^1\over\beta_0}
\biggl(
{\lambda\over 1-\lambda}
+\ln(1-\lambda)\biggr)
\ln\biggl(
{Q^2\over \mu_r^2}\biggr)
\nonumber \\
&&
+{4 A_q^1\beta_1\over\beta_0^3}
\biggl({1\over 2}
\ln^2
(1-\lambda)
+{\ln(1-\lambda)\over 1-\lambda}
+{\lambda\over 1-\lambda}\biggr)
\nonumber \\
\beta_1 &=&2\biggl({153-19 n_{f}\over 3}\biggr)
\nonumber\\
A_q^2&=&{C_F\over 2}
\biggl[{67\over 6}-{\pi^2\over 2}-{5\over 9}n_{f}\biggr]
\nonumber \\
{\overline B}_{q,N}^1&=& -{3\over 2}C_F+2\gamma^1_{qq,N}+A_q^1\ln\biggl({M_{VH}^2
\over Q^2}\biggr)
\, .
\end{eqnarray}
and $n_{f}$ is the number of light flavors.  The anomalous dimensions $\gamma_{ab,N}$ are the Mellin transforms of the DGLAP splitting functions, $P_{ab}$~\cite{Ellis:1991qj}:
\begin{equation}
\gamma_{ab,N}=\sum^{\infty}_{n=1}\left({\alpha_s\over \pi}\right)^n \gamma^n_{ab,N}\equiv\int^1_0dz z^{N-1}P_{ab}(z)
\end{equation}

The process dependence arises through
 $H_N^{VH(1)}$~\cite{Altarelli:1979ub,deFlorian:2001zd},
\begin{eqnarray}
H_{N,q\bar{q}\leftarrow qg}^{VH(1)}&=&\gamma^1_{qg,N} \log{Q^2\over \mu^2_f}+{1\over 2(N+1)(N+2)}\\
H_{N,q\bar{q}\leftarrow q\bar{q}}^{VH(1)}&=&
C_F\biggl(
{1\over N(N+1)}+
{\pi^2\over 6}\biggr)
+{1\over 2} A^{VH} 
\nonumber \\
&& -{C_F\over 2}\biggl[-3+
\log\biggl({M_{VH}^2\over Q^2}
\biggr)\biggr]\ln
\biggl(
{M_{VH}^2\over Q^2}\biggr)
+2 \gamma_{qq,N}^1\ln
\biggl({Q^2\over \mu_f^2}\biggr)\, ,
\end{eqnarray}
where,
\begin{eqnarray}
A^{VH}&=&C_F\biggl(-8+{2\pi^2\over 3}\biggr)\, .
\end{eqnarray}

\section*{Appendix B: Threshold Resummation}
In this appendix, we list the functions needed for the threshold 
resummation of Section \ref{sec_thresh}, taken 
from Ref.~\cite{Becher:2007ty}.  
All formulae in this appendix can be found in Ref. \cite{Becher:2007ty}, 
but we include them for the convenience of the reader.

The running kernel $U$ is defined as 
\begin{equation}
U(M,\mu_h,\mu_s,\mu_f)=\left(\frac{M^2}{\mu_h^2}\right)^{-2a_\Gamma(\mu_h,\mu_s)}
\exp\left[4S(\mu_h,\mu_s)-2a_{\gamma^V}(\mu_h,\mu_s)+4a_{\gamma^\phi}(\mu_s,\mu_f)\right],
\end{equation}  
where $a_\gamma$ is the anomalous exponent of $\gamma$ defined by
\begin{equation}
a_\gamma(\nu,\mu)=-\int_{\alpha_s(\nu)}^{\alpha_s(\mu)}d\alpha\frac{\gamma(\alpha)}{\beta(\alpha)},
\label{rgeA}
\end{equation}
and $S$  is the Sudakov exponent 
\begin{equation}
\label{rgeS}
S(\nu,\mu)=-\int_{\alpha_s(\nu)}^{\alpha_s(\mu)}d\alpha\frac{\Gamma_{\textrm{cusp}}(\alpha)}{\beta(\alpha)}
\int_{\alpha_s(\nu)}^\alpha\frac{d\alpha'}{\beta(\alpha')}.
\end{equation}

The renormalization group equations, Eqs.~(\ref{rgeA}) and (\ref{rgeS}), can  be solved perturbatively.
The anomalous dimensions are expanded as
\begin{equation}
\gamma(\alpha_s) = \gamma_0 \frac{\alpha_s}{4\pi} + \gamma_1\left(\frac{\alpha_s}{4\pi}\right)^2 
+ \gamma_2\left(\frac{\alpha_s}{4\pi}\right)^3 + \cdots
\end{equation}
The solutions to Eqs.~(\ref{rgeA}) and (\ref{rgeS}) are then
\begin{eqnarray}
a_\gamma(\nu,\mu) &=& \frac{\gamma_0}{2\beta_0}\left\{\ln\frac{\alpha_s(\mu)}{\alpha_s{\nu}} 
+ \left(\frac{\gamma_1}{\gamma_0} - \frac{\beta_1}{\beta_0}\right)\frac{\alpha_s(\mu) - \alpha_s(\nu)}{4\pi}\right.\nonumber\\
&&  \left. + \left[\frac{\gamma_2}{\gamma_0} - \frac{\beta_2}{\beta_0} - \frac{\beta_1}{\beta_0}\left(\frac{\gamma_1}{\gamma_0} - \frac{\beta_1}{\beta_0}\right)\right] \frac{\alpha_s^2(\mu) - \alpha_s^2(\nu)}{32\pi^2} + \cdots \right\},
\end{eqnarray}
and
\begin{eqnarray}
S(\nu,\mu) &=& \frac{\Gamma_0}{4\beta_0^2}\left\{\frac{4\pi}{\alpha_s(\nu)}\left(1 - \frac1r - \ln r\right) + \left(\frac{\Gamma_1}{\Gamma_0}-\frac{\beta_1}{\beta_0}\right)(1 - r + \ln r) + \frac{\beta_1}{2\beta_0} \ln^2 r\right.\nonumber\\
&& + \frac{\alpha_s(\nu)}{4\pi}\left[\left(\frac{\beta_1\Gamma_1}{\beta_0\Gamma_0} - \frac{\beta_2}{\beta_0}\right)( 1 - r + r \ln r)
+ \left(\frac{\beta_1^2}{\beta_0^2} - \frac{\beta_2}{\beta_0}\right)(1 - r)\ln r\right.\nonumber\\
&&\left.\phantom{+ \frac{\alpha_s(\nu)}{4\pi}} - \left(\frac{\beta_1^2}{\beta_0^2} - \frac{\beta_2}{\beta_0} - \frac{\beta_1\Gamma_1}{\beta_0\Gamma_0} + \frac{\Gamma_2}{\Gamma_0}\right) \frac{(1-r)^2}2\right]\nonumber\\
&&+\left(\frac{\alpha_s(\nu)}{4\pi}\right)^2\left[\left(\frac{\beta_1\beta_2}{\beta_0^2}-\frac{\beta_1^3}{2\beta_0^3}-\frac{\beta_3}{2\beta_0}
+\frac{\beta_1}{\beta_0}\left(\frac{\Gamma_2}{\Gamma_0}-\frac{\beta_2}{\beta_0}+\frac{\beta_1^2}{\beta_0^2}-\frac{\beta_1\Gamma_1}{\beta_0\Gamma_0}\right)\frac{r^2}{2}\right)\ln r\right. \nonumber\\
&&\phantom{+\left(\frac{\alpha_s(\nu)}{4\pi}\right)^2}+\left(\frac{\Gamma_3}{\Gamma_0}-\frac{\beta_3}{\beta_0}+\frac{2\beta_1\beta_2}{\beta_0^2}+\frac{\beta_1^2}{\beta_0^2}\left(\frac{\Gamma_1}{\Gamma_0}-\frac{\beta_1}{\beta_0}\right)-\frac{\beta_2\Gamma_1}{\beta_0\Gamma_0}-\frac{\beta_1\Gamma_2}{\beta_0\Gamma_0}\right)\frac{(1-r)^3}{3}
\nonumber\\
&&\phantom{+\left(\frac{\alpha_s(\nu)}{4\pi}\right)^2}+\left(\frac{3\beta_3}{4\beta_0}-\frac{\Gamma_3}{2\Gamma_0}+\frac{\beta_1^3}{\beta_0^3}-\frac{3\beta_1^2\Gamma_1}{4\beta_0^2\Gamma_0}
+\frac{\beta_2\Gamma_1}{\beta_0\Gamma_0}+\frac{\beta_1\Gamma_2}{4\beta_0\Gamma_0}-\frac{7\beta_1\beta_2}{4\beta_0^2}\right)(1-r)^2
\nonumber\\
&&\left.\left.\phantom{+\left(\frac{\alpha_s(\nu)}{4\pi}\right)^2}+\left(\frac{\beta_1\beta_2}{\beta_0^2}-\frac{\beta_3}{\beta_0}-\frac{\beta_1^2\Gamma_1}{\beta_0^2\Gamma_0}+\frac{\beta_1\Gamma_2}{\beta_0\Gamma_0}\right)\frac{1-r}{2}\right]+\cdots\right\}\, ,
\end{eqnarray}
where $r\equiv \alpha_s(\mu)/\alpha_s(\nu)$.

The cusp anomalous dimension is known to three-loops \cite{Korchemskaya:1992je,Moch:2004pa}. The coefficients are
\begin{eqnarray}
\Gamma_0 &=& 4C_F, \nonumber\\
\Gamma_1 &=& 4 C_F\left[\left(\frac{67}9 - \frac{\pi^2}3 \right) C_A - \frac{20}9 T_F n_f\right],\nonumber\\
\Gamma_2 &=& 4 C_F\left[C_A^2\left(\frac{245}6 - \frac{134\pi^2}{27} + \frac{11 \pi^4}{45} + \frac{22}3\zeta_3 \right)
+ C_A T_F n_f  \left(-\frac{418}{27} + \frac{40\pi^2}{27} - \frac{56}3 \zeta_3\right)\right. \nonumber\\
&&\left.+ C_F T_F n_f\left(-\frac{55}3 + 16 \zeta_3\right) - \frac{16}{27} T_F^2 n_f^2\right].
\end{eqnarray}
The four-loop coefficient $\Gamma_3$ has not yet been calculated, so we use the Pad\'e approximate $\Gamma_3 = \Gamma_2^2/\Gamma_1$.  
The anomalous dimension $\gamma^V$ can be obtained from the partial three-loop on-shell quark form factor \cite{Moch:2005id}.  The coefficients are
\begin{eqnarray}
\gamma^V_0 &=& -6 C_F,\nonumber\\
\gamma^V_1 &=& C_F^2(-3 + 4\pi^2 - 48 \zeta_3) + C_F C_A \left(-\frac{961}{27} - \frac{11\pi^2}3 + 52 \zeta_3\right) + C_F T_F n_f\left(\frac{260}{27} + \frac{4\pi^2}3\right),\nonumber\\
\gamma^V_2 &=& C_F^3\left(-29 - 6\pi^2 - \frac{16\pi^4}5 - 136\zeta_3 + \frac{32\pi^2}3\zeta_3 + 480 \zeta_5\right)\nonumber\\
&& + C_F^2 C_A\left(-\frac{151}2 + \frac{410 \pi^2}9 + \frac{494\pi^4}{135} - \frac{1688}3\zeta_3 - \frac{16\pi^2}3 \zeta_3 - 240 \zeta_5\right)\nonumber\\
&& + C_F C_A^2\left(-\frac{139345}{1458} - \frac{7163\pi^2}{243} - \frac{83\pi^4}{45} + \frac{7052}9 \zeta_3 - \frac{88\pi^2}9 \zeta_3 - 272 \zeta_5\right)\nonumber\\
&&+ C_F^2 T_F n_f\left(\frac{5906}{27} - \frac{52 \pi^2}9 - \frac{56 \pi^4}{27} + \frac{1024}9 \zeta_3\right)\nonumber\\
&& + C_F C_A T_F n_f\left(-\frac{34636}{729} + \frac{5188\pi^2}{243} + \frac{44\pi^4}{45} - \frac{3856}{27} \zeta_3\right)\nonumber\\
&& + C_F T_F^2 n_f^2\left(\frac{19336}{729} - \frac{80 \pi^2}{27} - \frac{64}{27} \zeta_3\right).
\end{eqnarray}
The final anomalous dimension, $\gamma^\phi$, is known from the NNLO calculation of the Altarelli-Parisi splitting function \cite{Moch:2004pa}.  The coefficients are
\begin{eqnarray}
\gamma^\phi_0 &=& 3 C_F,\nonumber\\
\gamma^\phi_1 &=& C_F^2\left(\frac32 - 2\pi^2 + 24 \zeta_3\right) + C_F C_A \left(\frac{17}{6} + \frac{22\pi^2}9 - 12 \zeta_3\right) - C_F T_F n_f\left(\frac{2}{3} + \frac{8\pi^2}9\right), \nonumber\\
\gamma^\phi_2 &=& C_F^3\left(\frac{29}2 + 3\pi^2 + \frac{8\pi^4}5 + 68\zeta_3 - \frac{16\pi^2}3\zeta_3 -240 \zeta_5\right)\nonumber\\
&& + C_F^2 C_A\left(\frac{151}4 - \frac{205 \pi^2}9 - \frac{247\pi^4}{135} + \frac{844}3\zeta_3 + \frac{8\pi^2}3 \zeta_3 + 120 \zeta_5\right)\nonumber\\
&& + C_F C_A^2\left(-\frac{1657}{36} + \frac{2248\pi^2}{81} - \frac{\pi^4}{18} - \frac{1552}9 \zeta_3 + 40 \zeta_5\right)\nonumber\\
&&+ C_F^2 T_F n_f\left(-46 + \frac{20 \pi^2}9 + \frac{116 \pi^4}{135} - \frac{272}3 \zeta_3\right)\nonumber\\
&& + C_F C_A T_F n_f\left(40 -  \frac{1336\pi^2}{81} + \frac{2\pi^4}{45} + \frac{400}{9} \zeta_3\right)\nonumber\\
&& + C_F T_F^2 n_f^2\left(-\frac{68}{9} + \frac{160 \pi^2}{81} - \frac{64}{9} \zeta_3\right).
\end{eqnarray}

The other functions needed are the Wilson coefficient $C_V$ and the soft function $\tilde s_{\rm DY}$.  
The Wilson coefficient $C_V$ has the expansion,
\begin{equation}
C_V(-M^2 - i\epsilon,\mu) = 1 + \frac{C_F\alpha_s}{4\pi}\left(-L^2 + 3 L - 8 + \frac{\pi^2}6\right) 
  + C_F\left(\frac{\alpha_s}{4\pi}\right)^2(C_F H_F + C_A H_A + T_F n_f H_f),
\end{equation}
where $L = \ln(M^2/\mu^2) - i\pi$, and 
\begin{eqnarray}
H_F &=& \frac{L^4}2 - 3 L^3 + \left(\frac{25}2 - \frac{\pi^2}6 \right)L^2 + \left(-\frac{45}2 - \frac{3\pi^2}2 + 24\zeta_3\right)L
+\frac{255}8 + \frac{7\pi^2}2 - \frac{83\pi^4}{360} - 30\zeta_3,\nonumber\\
H_A &=& \frac{11}9 L^3 + \left(-\frac{233}{18} + \frac{\pi^2}3\right) L^2 + \left(\frac{2545}{54} + \frac{11\pi^2}9 - 26\zeta_3\right)L\nonumber\\
&&-\frac{51157}{648} - \frac{337\pi^2}{108} + \frac{11\pi^4}{45} + \frac{313}9\zeta_3,\nonumber\\
H_f &=& -\frac49 L^3 + \frac{38}9 L^2 + \left( -\frac{418}{27} - \frac{4\pi^2}9 \right) L + \frac{4085}{162} + \frac{23\pi^2}{27} + \frac49 \zeta_3. 
\end{eqnarray}
This agrees with the corresponding expression in \cite{Idilbi:2006dg}.

The soft function to two-loops is
\begin{equation}
\tilde s_{\rm DY}(\ell,\mu) = 1 + \frac{C_F\alpha_s}{4\pi}\left(2\ell^2 + \frac{\pi^2}3\right) + C_F\left(\frac{\alpha_s}{4\pi}\right)^2(C_F W_F + C_A W_A + T_F n_f W_f),
\end{equation}
where
\begin{eqnarray}
W_F &=& 2\ell^4 + \frac{2\pi^2}3\ell^2 + \frac{\pi^4}{18},\nonumber\\
W_A &=& -\frac{22}9 \ell^3 + \left(\frac{134}9 - \frac{2\pi^2}3\right)\ell^2 + \left(-\frac{808}{27} + 28\zeta_3\right) \ell
  + \frac{2428}{81} + \frac{67\pi^2}{54} -\frac{\pi^4}3 - \frac{22}9\zeta_3,\nonumber\\
W_f &=& \frac89 \ell^3 - \frac{40}9 \ell^2 + \frac{224}{27} \ell - \frac{656}{81} - \frac{10\pi^2}{27} + \frac89 \zeta_3.
\end{eqnarray}
This again agrees with the moment-space expression in \cite{Idilbi:2006dg}.

\section*{Acknowledgements}
The work of S.D.~and I.L.~is supported by the U.S. Department of Energy under grant
No.~DE-AC02-98CH10886. 
The work of T.H.~is supported in part by the US Department of Energy
under grant No.~DE-FG02-12ER41832, in part by PITT PACC.
The work of A.K.L.~and W.K.L.~is supported in part by the 
National Science Foundation under Grant No.~PHY-0854782.

\bibliography{outline}
\end{document}